\documentclass[pra,twocolumn,superscriptaddress,amsmath,amssymb,floatfi,noshowpacs]{revtex4}
\usepackage{graphicx}
\usepackage{amsmath}

\begin{document}
\title{\textbf{Collective motion of an atom array under laser illumination}}
\author{Ephraim Shahmoon}
\affiliation{Department of Physics, Harvard University, Cambridge, Massachusetts 02138, USA}
\author{Mikhail D.~Lukin}
\affiliation{Department of Physics, Harvard University, Cambridge, Massachusetts 02138, USA}
\author{Susanne F.~Yelin}
\affiliation{Department of Physics, Harvard University, Cambridge, Massachusetts 02138, USA}
\affiliation{Department of Physics, University of Connecticut, Storrs, Connecticut 06269, USA}
\date{\today}

\begin{abstract}
We develop a theoretical formalism for the study of light-induced motion of atoms trapped in a two-dimensional (2D) array, considering the effect of multiple scattering of light between the atoms.
We find that the atomic motion can be described by a collective diffusion equation, wherein laser-induced dipole-dipole forces couple the motion of different atoms.
This coupling leads to the formation of collective mechanical modes of the array atoms, whose spatial structure and stability depend on the parameters of the illuminating laser and the geometry of the 2D array.
We demonstrate the application of our formalism for the analysis of light-induced heating of the 2D array. The presented approach should be useful for treating the optomechanical properties of recently proposed quantum optical platforms made of atomic arrays.
\end{abstract}

\pacs{} \maketitle

\section{Introduction}
The study of the motion of atoms under the influence of light has lead to major breakthroughs in modern atomic and optical physics. The many applications of laser-induced cooling and trapping of atoms \cite{CCT,CCTb} include the creation and study of Bose-Einstein condensation of atoms \cite{BEC1,BEC2,BEC3,PS}, the exploration of interacting quantum gases and many-body systems \cite{OL,QS,TH,MBL}, and atom-based quantum optics and information \cite{QO}.

The standard description of atomic motion under laser illumination treats each atom separately, by considering only the individual-atom response for light \cite{CCT}. Therefore, effects related to the scattering of light between different atoms, that is, dipole-dipole interactions, are typically neglected. Such a treatment is valid in cases where far-off-resonant light is used, such as in the trapping of atoms in optical lattices. For a dense enough atomic media, however, and in particular when resonant light is shined, collective dipolar effects become increasingly important \cite{RUO,RUO2}.

An interesting arena for the exploration of the role of collective dipolar effects is that of \emph{ordered} atomic arrays. Recent studies have shown how collective dipolar effects can be harnessed for the design of non-trivial responses of atomic arrays to light \cite{coop,ADM,janos,ADM2,janos2,CHA,HEN,ABA,ANA}. The main idea is that dipolar interactions between the atoms lead to the formation of collective dipole excitations supported by the atom array. The resonant frequencies of these collective dipolar modes, are shifted from the ``bare", single-atom resonance, due to the dipole-dipole interactions.
For light at cooperative resonance, namely, at resonance with the collective dipole modes, it was found that a two-dimensional (2D) atomic array can act as a perfect mirror \cite{coop,ADM}, couple collimated light to a single atom \cite{coop}, support topological photonic modes \cite{janos,ADM2,janos2}, and enhance quantum memories and clocks \cite{CHA,HEN}. All of these previous works rely on the collective response of the \emph{internal} (dipolar) states of the atoms, and their influence on the propagation of light. However, dipolar interactions can also lead to light-induced collective effects in the external, \emph{motional} degrees of freedom of the atoms \cite{GIO,OD2,CHA1,LIDDIna,RIT,RIT2,ESS,LEO}, which were not considered for such atomic arrays.

The purpose of the present study is to develop a quantum-mechanical formalism for the description of collective mechanical effects of 2D atomic arrays under laser illumination. Specifically, we consider the longitudinal motion of atoms trapped in a 2D optical lattice, under the influence of a continuous illumination at normal incidence (Fig. 1a,b), and neglecting saturation of the atoms. We find that the dynamics of the atoms are governed by a \emph{collective diffusion equation}, wherein atoms with a renormalized internal-state response (cooperative resonance) are motionally coupled via laser-induced dipole-dipole forces (Fig. 1c). These laser-induced interactions then lead to the formation of collective mechanical modes of the atoms, whose spectrum and spatial structure are determined by the parameters of the laser and the array. Some of the peculiar collective mechanical effects we find are the possibility for gapped or unstable mechanical modes. The analysis is applied to study the thermalization and heating of an atom array under illumination.

The formalism developed here can be seen as a generalization of the single-atom theory of Ref. \cite{CCT} to the quantum-mechanical treatment of the motion of an atomic array, wherein multiple scattering of light between the atoms is significant. Our results are useful for the analysis of the influence of light-induced motion, on the quantum optical applications of atom arrays studied previously \cite{coop,janos,ADM2,janos2,ABA,CHA,ANA,HEN,ADM}, and even more so, to the exploration of new opportunities for optomechanics using ordered atomic arrays.

\emph{Outline.---} The main general result, of a collective diffusion equation for the array atoms, is presented and explained in Sec. IV D; readers not interested in its derivation may skip the preceding sections. The rest of the paper is organized as follows. In Sec. II, we present the model and derive general equations of motion for the internal and external degrees of freedom of a general collection of atoms, and without neglecting their photon-mediated interactions. In Secs. III and IV we further consider our main assumptions of small-amplitude motion around a 2D array geometry and the separation of internal-external time-scales, respectively, arriving at the collective diffusion equation (\ref{90}).
Section V is devoted to the analysis of the resulting collective mechanical modes of the array. As an application of our formalism, we study in Sec. VI, the durability of the atom array under illumination, by considering the resulting heating of the array atoms. Our conclusions are finally presented in Sec. VII.

\begin{figure}
\begin{center}
\includegraphics[width=\columnwidth]{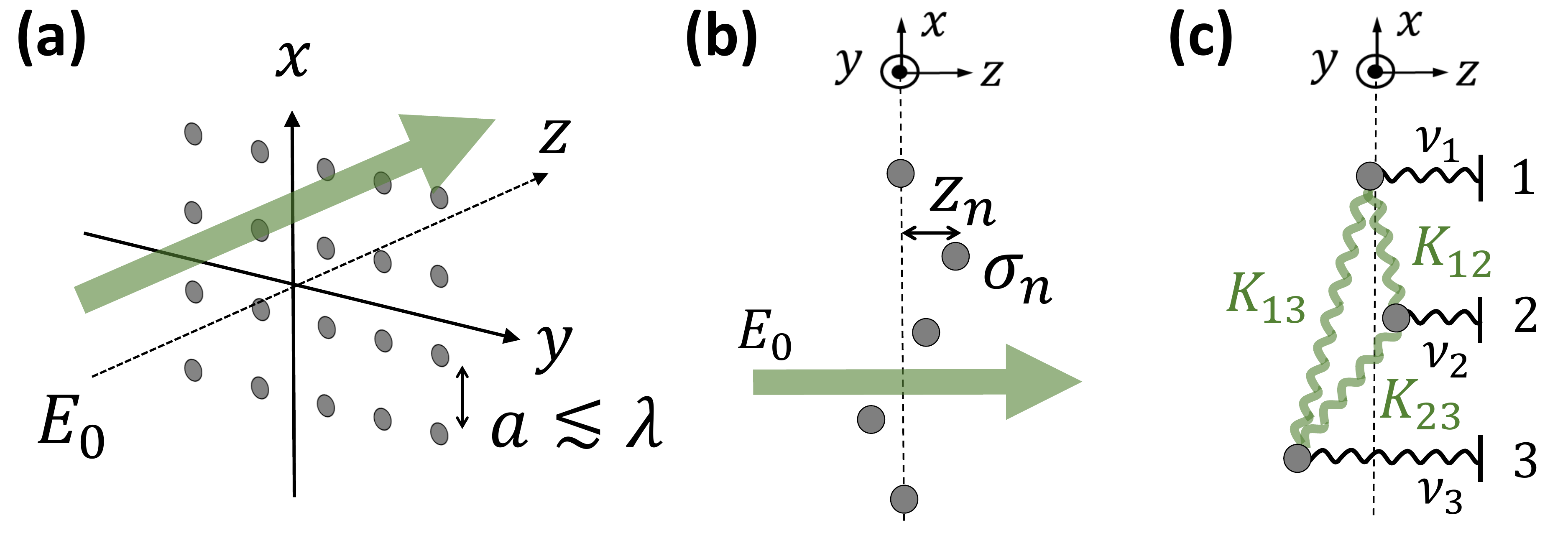}
\caption{\small{
Light-induced motion of atoms trapped in a 2D array. (a) Laser field $E_0$ propagating along the longitudinal $z$-axis illuminates atoms trapped in an optical lattice spanning the $xy$ plane, with a lattice constant $a\lesssim \lambda$, $\lambda$ being the transition wavelength of the atoms. (b) The dynamics of an atom $n$ is characterized by its internal-state, $\hat{\sigma}_n$, and its longitudinal motion inside the trap, $\hat{z}_n$ (assuming tight  trapping, and hence fixed positions, along $xy$). (c) Laser-induced dipole-dipole forces couple the motion of different atoms (``spring constant" $K_{nm}$). Together with the optical-lattice traps of individual atoms (black ``spring constants" $\nu_n$), the resulting spring model in the figure describes the conservative part of the dynamics captured by Eq. (\ref{90}).
 }} \label{fig1}
\end{center}
\end{figure}

\section{Atomic equations of motion}

The main result of this section is the coupled Heisenberg-Langevin equations for the internal and external atomic degrees of freedom, Eq. (\ref{HL}). Starting with the general atom-photon Hamiltonian and eliminating the photon modes (Markov approximation), no assumptions are made at this stage, on the polarization or propagation direction of the exciting continuous-wave laser. The role of dipole-dipole interactions and forces, Eqs. (\ref{Dnm}) and (\ref{Anm}), is emphasized.

\emph{System and Hamiltonian.---} Consider $N$ identical atoms with a $J=0$ to $J=1$ transition, such that each atom has a ground state $|g\rangle$ and three degenerate excited states $|e_i\rangle$ (with $i=x,y,z$ the polarization axes). The atoms form a 2D lattice in the $xy$ plane, in which they are tightly trapped and assumed motionless along the $xy$ directions. The trapping along the longitudinal $z$-axis is taken to be finite however, with a trap frequency $\nu_n$ and longitudinal coordinate $\hat{z}_n$ and momentum $\hat{p}_n$ for an atom $n=1,..,N$.  The atoms are illuminated by a continuous laser at a frequency $\omega_L$ and field amplitude $\mathbf{E}_0(\mathbf{r})e^{-i\omega_L t}=\sum'_{\mathbf{q}} e^{i\mathbf{q}\cdot\mathbf{r}}e^{-i\omega_L t}\mathbf{E}_{0,\mathbf{q}}$, where the notation $\sum'_{\mathbf{q}}$ means that the sum on the spatial Fourier components, $\mathbf{E}_{0,\mathbf{q}}$, is constrained by $|\mathbf{q}|=q=\omega_L/c$. The Hamiltonian in the rotating-wave approximation is 
\begin{eqnarray}
H&=&H_A+H_M+H_V+H_{ALM}(t)+H_{AVM},
\nonumber\\
H_A&=&\sum_{n=1}^N\sum_{i=x,y,z}\hbar\omega_a\hat{\sigma}_{ni}^{\dag}\hat{\sigma}_{ni},
\nonumber\\
H_M&=&\sum_n\left[\frac{\hat{p}_n^2}{2m}+\frac{1}{2}m\nu_n^2\hat{z}_n^2\right],
\nonumber\\
H_V&=&\sum_{\mathbf{k}\mu}\hbar\omega_{\mathbf{k}}\hat{a}^{\dag}_{\mathbf{k}\mu}\hat{a}_{\mathbf{k}\mu}, \quad \omega_{\mathbf{k}}=|\mathbf{k}|c,
\nonumber\\
H_{ALM}&=&-\hbar\sum_{ni} \sideset{}{'} \sum_{\mathbf{q}}\left[\Omega_{\mathbf{q}}^i e^{i\mathbf{q}\cdot \hat{\mathbf{r}}_n}e^{-i\omega_L t}\hat{\sigma}_{ni}^{\dag}+\mathrm{h.c.}\right],
\nonumber\\
H_{AVM}&=&-\hbar\sum_{ni}\sum_{\mathbf{k}\mu}\left[ig_{\mathbf{k}\mu}^i e^{i\mathbf{k}\cdot\hat{\mathbf{r}}_n}\hat{\sigma}_{ni}^{\dag}\hat{a}_{\mathbf{k}\mu}+\mathrm{h.c.}\right].
\label{H}
\end{eqnarray}
Here $\hat{\sigma}_{ni}=|g\rangle_n\langle e_i|$ is the lowering operator of the $i=x,y,z$ atomic transition, $|g\rangle\rightarrow |e\rangle_i$, of an atom $n$ with transition energy $\omega_a$.
$\Omega_{\mathbf{q}}^i=d\mathbf{e}_i\cdot \mathbf{E}_{0,\mathbf{q}}/\hbar$ is the $\mathbf{q}$ spatial component of the Rabi frequency acting on the $i$ atomic transition with dipole matrix element $d$ (taken identical for all $i=x,y,z$). The atomic coordinate, $\hat{\mathbf{r}}_n=(\mathbf{r}^{\bot}_n,\hat{z}_n)$, appears as an operator via its longitudinal component $\hat{z}_n$, whereas the in-plane positions $\mathbf{r}^{\bot}_n$ are fixed. The bosonic operators $\hat{a}_{\mathbf{k}\mu}$ describe the photon vacuum modes with wavevector $\mathbf{k}$ and polarization $\mu$, whose dipole couplings to the atomic transition $i=x,y,z$ is given by $g_{\mathbf{k}\mu}^i=\sqrt{\omega_{\mathbf{k}}/(2\varepsilon_0\hbar V)}d\mathbf{e}_{\mu}\cdot\mathbf{e}_i$, with $V$ the free-space quantization volume.

\emph{Vacuum field as a reservoir.---} Moving to the laser-rotated picture and using standard methods, we obtain the Heisenberg-Langevin equations of motion for the atomic operators. This is performed as usual by the Markov approximation: taking the assumption, $\tau_s\gg 1/\omega_L, c/L$, with $\tau_s$ being a typical time-scale for the evolution of atomic degrees of freedom, and $L$ being the linear extent of the atomic system, the photon vacuum variables can be eliminated and the resulting equations of motion for the atomic variables read (Appendix A)
\begin{eqnarray}
\dot{\tilde{\sigma}}_{ni}&=&(i\delta_L-\gamma/2)\tilde{\sigma}_{ni}+\sum_j(\tilde{\sigma}_{ni}\tilde{\sigma}_{ni}^{\dag}\delta_{ij}-\tilde{\sigma}_{nj}^{\dag}\tilde{\sigma}_{ni})
\nonumber\\
&\times&
\left[i \sideset{}{'}\sum_{\mathbf{q}}e^{i\mathbf{q}\cdot \hat{\mathbf{r}}_n}\Omega_{\mathbf{q}}^j +i\sum_{k_z}e^{ik_z \hat{z}_n}\delta\hat{\Omega}_{k_z,n}^j(t) \right.
\nonumber \\
&& \left.-\sum_{m\neq n}\sum_{l}\mathcal{D}_{jl}(\hat{\mathbf{r}}_n-\hat{\mathbf{r}}_m)\tilde{\sigma}_{ml}\right],
\nonumber\\
\dot{\hat{p}}_n&=&-m\nu_n^2\hat{z}_n+\sum_i \sideset{}{'}\sum_{\mathbf{q}}\hbar q_z\left[i \tilde{\sigma}_{ni}^{\dag} e^{i\mathbf{q}\cdot\hat{\mathbf{r}}_n}\Omega_{\mathbf{q}}^i +\mathrm{h.c.}\right]
\nonumber\\
&&+\sum_i\sum_{k_z}\hbar k_z \left[i \tilde{\sigma}_{ni}^{\dag} e^{ik_z \hat{z}_n}\delta\hat{\Omega}_{k_z,n}^i(t)+\mathrm{h.c.}\right]
\nonumber\\
&&-\sum_{m\neq n} \sum_{i,j}\left[\tilde{\sigma}_{ni}^{\dag}A_{ij}(\hat{\mathbf{r}}_n-\hat{\mathbf{r}}_m)\tilde{\sigma}_{mj}+\mathrm{h.c.}\right],
\nonumber\\
\dot{\hat{z}}_n&=&\hat{p}_n/m.
\label{HL}
\end{eqnarray}
Here $\tilde{\sigma}_{ni}=\hat{\sigma}_{ni} e^{i\omega_L t}$ is the slow envelope of the internal dynamics, $\delta_L=\omega_L-\omega_a$ is the laser detuning (with the Lamb shift absorbed into $\omega_a$), and $\gamma=d^2\omega_L^3/(3\pi \varepsilon_0\hbar c^3)$ is the spontaneous emission rate evaluated at $\omega_L$. The corresponding Langevin noise due to the photon vacuum is given by $i\sum_{k_z}e^{ik_z\hat{z}_n}\delta\hat{\Omega}_{k_z,n}^i(t)$, with
\begin{equation}
\delta\hat{\Omega}_{k_z,n}^i(t)=\sum_{\mathbf{k}_{\bot}\mu} i g_{\mathbf{k}\mu}^i e^{i\mathbf{k}_{\bot}\cdot\mathbf{r}^{\bot}_n}e^{-i(\omega_{\mathbf{k}}-\omega_L)t}\hat{a}_{\mathbf{k}\mu}, \quad \mathbf{k}=(\mathbf{k}_{\bot},k_z).
\label{F}
\end{equation}
Here $\delta\hat{\Omega}_{k_z,n}^i(t)$ is the $k_z$ component of the plane-wave expansion of the Langevin noise, where $\hat{a}_{\mathbf{k}\mu}=\hat{a}_{\mathbf{k}\mu}(t=0)$. The force due to photon recoil from the laser and vacuum appears in the equation for $\hat{p}_n$ via the photon momenta along $z$, $\hbar q_z$ and  $\hbar k_z$, respectively, with $q_z=\mathbf{e}_z\cdot\mathbf{q}$ and $k_z=\mathbf{e}_z\cdot\mathbf{k}$.

\emph{Dipole-dipole interactions.---} The equation for $\tilde{\sigma}_{ni}$ includes the interaction term with all other atoms $m$ via the complex interaction kernel,
\begin{equation}
\mathcal{D}_{ij}(\hat{\mathbf{r}}_n-\hat{\mathbf{r}}_m)=-i\frac{3}{2}\gamma\lambda G_{ij}(\omega_L,\hat{\mathbf{r}}_n-\hat{\mathbf{r}}_m)\equiv\frac{1}{2}\Gamma^{ij}_{nm}+i\Delta^{ij}_{nm},
\label{Dnm}
\end{equation}
where $\lambda=2\pi c/\omega_L$ is the laser's wavelength. Here $G_{ij}(\omega,\mathbf{r})$ is the dyadic Green's function tensor of the electromagnetic field in free space \cite{NH} [Eq. (\ref{G}), Appendix A], where $\Gamma^{ij}_{nm}$ and $\Delta^{ij}_{nm}$ are related to its imaginary (cooperative dissipation) and real (dipole-dipole interaction) parts, respectively. The appearance of the dipole-dipole interaction term is analogous to the multiple scattering of electromagnetic fields between the atoms, which lead to cooperative scattering phenomena.
Dipolar interactions lead additionally to the force term in the equation for $\hat{p}_n$ with the interaction kernel
\begin{equation}
A_{ij}(\hat{\mathbf{r}}_n-\hat{\mathbf{r}}_m)=-i\hbar\left.\frac{\partial}{\partial z}\mathcal{D}_{ij}(\mathbf{r})\right|_{\mathbf{r}=\hat{\mathbf{r}}_n-\hat{\mathbf{r}}_m}.
\label{Anm}
\end{equation}
This term describes the force acting on atom $n$ due to the pressure impinged by laser photons scattered off by the atom $m$. This will become clearer in the following, when we relate this force to the laser-induced dipole-dipole interaction potential \cite{THI,SAL,LIDDI} built between the atoms $n$ and $m$.
The explicit expressions for both $G_{ij}(\omega,\mathbf{r})$ and $A_{ij}(\mathbf{r})$ are given in Appendix A.

\section{Small-amplitude motion}

In this section we exploit the ordered lattice structure of the atomic array, in order to simplify the atomic equations of motions. The main assumption taken here
is of atomic motion that is sufficiently small for the 2D lattice structure to stay roughly intact. More precisely, we assume small differences in the longitudinal translations of different atoms, as expressed in Eq. (\ref{46}) below. This assumption allows to exploit the discrete translational symmetry of the 2D lattice.
For simplicity, we further assume two-level atoms polarized along the 2D lattice plane ($xy$), illuminated by weak (non-saturating) paraxial illumination along $z$. The main result is the atomic equations of motions, Eqs. (\ref{60}) and (\ref{62}).

\subsection{The small-amplitude assumption}
For the consideration of small-amplitude motion around the 2D lattice positions $\mathbf{r}^{\bot}_n$ at $z=0$ , we assume small differences in atomic positions,
\begin{eqnarray}
|\hat{z}_n-\hat{z}_m|\ll \lambda \quad \Leftrightarrow \quad q |\hat{z}_n-\hat{z}_m|\ll 1.
\label{46}
\end{eqnarray}
The validity of the assumption (\ref{46}) should hold throughout the dynamics described below. This assumption allows to consider Eqs. (\ref{HL}) to linear order in $q |\hat{z}_n-\hat{z}_m|$, which affects the dipole-dipole kernels as follows. The force kernel $A_{ij}(\hat{\mathbf{r}}_n-\hat{\mathbf{r}}_m)$ has a vanishing zeroth order and we obtain [see Eqs. (\ref{AF}) and (\ref{F})],
\begin{eqnarray}
A_{ij}(\hat{\mathbf{r}}_n-\hat{\mathbf{r}}_m)\approx A'^{ij}_{nm}q(\hat{z}_n-\hat{z}_m), \quad A'^{ij}_{nm}=-\frac{3}{4}\hbar q \gamma F_{nm}^{ij},
\nonumber\\
\label{47a}
\end{eqnarray}
with $F_{nm}^{ij}$ given in Appendix A [Eqs. (\ref{AF}) and (\ref{F})]. For the interaction kernel, $\mathcal{D}_{ij}(\hat{\mathbf{r}}_n-\hat{\mathbf{r}}_m)$, the first order vanishes and we retain only the zeroth order,
\begin{equation}
\mathcal{D}_{ij}(\hat{\mathbf{r}}_n-\hat{\mathbf{r}}_m)\approx \mathcal{D}_{ij}(\mathbf{r}^{\bot}_{n}-\mathbf{r}^{\bot}_{m})\equiv \mathcal{D}^{ij}_{nm}.
\label{47b}
\end{equation}

\subsection{Non-saturated two-level atoms}
For simplicity, we consider from now on, two-level atoms with single relevant transition $|g\rangle \rightarrow |e\rangle$ whose dipolar matrix element $\mathbf{d}=d\mathbf{e}_d$ is parallel to the $xy$ plane, $\mathbf{e}_d\bot\mathbf{e}_z$. Then, we can drop the indices $i,j$ and any summation over them,
\begin{equation}
\mathcal{D}^{ij}_{nm} \rightarrow \mathcal{D}_{nm}, \quad  A'^{ij}_{nm} \rightarrow A'_{nm}, \quad \tilde{\sigma}_{ni}\rightarrow \tilde{\sigma}_{n}, \quad \Omega^i\rightarrow \Omega,
\label{50}
\end{equation}
where the scalar quantities imply the projection on the dipole orientation $\mathbf{e}_d$ (e.g. $\mathcal{D}_{nm}=\mathbf{e}_d^{\dag}\cdot \overline{\overline{\mathcal{D}}}_{nm}\cdot\mathbf{e}_d$ for a tensor $\mathcal{D}^{ij}_{nm}=\mathbf{e}_i^{\dag}\cdot \overline{\overline{\mathcal{D}}}_{nm}\cdot\mathbf{e}_j$).
We further assume that the illuminating laser is weak enough so that the atoms are far from saturation,
leading to linearly-responding atomic dipoles, $(\tilde{\sigma}_{ni}\tilde{\sigma}_{ni}^{\dag}\delta_{ij}-\tilde{\sigma}_{nj}^{\dag}\tilde{\sigma}_{ni})\approx \delta_{ij}\rightarrow 1$ [see corresponding condition in (\ref{48}) below].
Applying all the above to the equation for $\tilde{\sigma}_n$ in (\ref{HL}), we obtain,
\begin{eqnarray}
\dot{\tilde{\sigma}}_{n}&=&(i\delta_L-\gamma/2)\tilde{\sigma}_{n}-\sum_{m\neq n}\mathcal{D}_{nm}\tilde{\sigma}_{m}
\nonumber \\
&+&i \sideset{}{'}\sum_{\mathbf{q}}e^{i\mathbf{q}\cdot \hat{\mathbf{r}}_n}\Omega_{\mathbf{q}} +i\sum_{k_z}e^{ik_z \hat{z}_n}\delta\hat{\Omega}_{k_z,n}(t).
\label{51a}
\end{eqnarray}

The above equation can be transformed to 2D lattice-wavevector space, $\mathbf{k}_{\bot}=(k_x,k_y)$, via $\frac{1}{N}\sum_{n=1}^N e^{-i\mathbf{k}_{\bot}\cdot\mathbf{r}^{\bot}_n}(...)$, yielding
\begin{eqnarray}
&&\dot{\tilde{\sigma}}_{\mathbf{k}_{\bot}}=(i\delta_L-\gamma/2)\tilde{\sigma}_{\mathbf{k}_{\bot}}-\mathcal{D}_{\mathbf{k}_{\bot}}\tilde{\sigma}_{\mathbf{k}_{\bot}}
\nonumber \\
&&+\frac{1}{N}\sum_n e^{-i\mathbf{k}_{\bot}\cdot\mathbf{r}^{\bot}_n} \left[i \sideset{}{'}\sum_{\mathbf{q}}e^{i\mathbf{q}\cdot \hat{\mathbf{r}}_n}\Omega_{\mathbf{q}} +i\sum_{k_z}e^{ik_z\hat{z}_n}\delta\hat{\Omega}_{k_z,n}(t)\right],
\nonumber\\
\label{53}
\end{eqnarray}
with $\mathcal{D}_{\mathbf{k}_{\bot}}=\sum_{n\neq 0}e^{-i\mathbf{k}_{\bot}\cdot \mathbf{r}_{n0}^{\bot}}\mathcal{D}_{0n}$ and $\tilde{\sigma}_{\mathbf{k}_{\bot}}=\frac{1}{N}\sum_{n=1}^N\tilde{\sigma}_{n}e^{-i\mathbf{k}_{\bot}\cdot\mathbf{r}^{\bot}_n}$.
The wavevector $\mathbf{k}_{\bot}$  lies within the first Brillouin zone of the reciprocal 2D lattice (e.g. $k_{x,y}\in [-\pi/a,\pi/a]$ for a square lattice with spacing $a$). The discrete translational symmetry of an infinite lattice (practically, $\sqrt{N}\gg 1$, see below) was used here to find $\frac{1}{N}\sum_n\sum_{m\neq n}e^{-i\mathbf{k}_{\bot}\cdot \mathbf{r}_n^{\bot}}\mathcal{D}_{nm} e^{i\mathbf{k}'_{\bot}\cdot \mathbf{r}_m^{\bot}}=\mathcal{D}_{\mathbf{k}_{\bot}}\delta_{\mathbf{k}_{\bot},\mathbf{k}'_{\bot}}$.

\subsection{Paraxial illumination}
For the strongest mechanical effect we also assume that the incident beam propagates perpendicular to the array along $z$. More precisely, we assume that the beam is paraxial in the sense that, $|\mathbf{q}_{\bot}|\ll |q_z|\approx q=\omega_L/c$, and approximate
\begin{eqnarray}
&&\sideset{}{'}\sum_{\mathbf{q}}e^{i\mathbf{q}\cdot \hat{\mathbf{r}}_n}\Omega_{\mathbf{q}}\approx \sum_{s=\pm}\Omega_{ns} e^{isq\hat{z}_n},
\nonumber\\
&&\Omega_{ns}=\sum_{\mathbf{q}_{\bot}}e^{i\mathbf{q}_{\bot}\cdot \hat{\mathbf{r}}^{\bot}_n}\Omega_{\mathbf{q}_{\bot},q_z=s|q_z|}
\label{59}
\end{eqnarray}
where $s=\pm$ denotes right/left-propagating laser. Since the laser drives all excitations and motion in the system, the relevant atomic dipole modes, $\tilde{\sigma}_{\mathbf{k}_{\bot}}$, and the relevant vacuum modes, $\hat{a}_{\mathbf{k}\mu}$, are also paraxial. We then use $|k_z|\approx k=\omega_{\mathbf{k}}/c$ for the vacuum fields, and
\begin{equation}
\mathcal{D}_{\mathbf{k}_{\bot}}\approx \mathcal{D}_{\mathbf{k}_{\bot}=0}=\frac{\Gamma}{2}+i\Delta,
\label{Dk}
\end{equation}
for the dipole-dipole interaction kernel. The interaction-induced cooperative shift $\Delta$ and width $\Gamma$, are given by [c.f. Eq. (\ref{Dnm})]
\begin{eqnarray}
\Delta&=&-\frac{3}{2}\gamma\lambda\sum_{n\neq 0} \mathrm{Re}[\mathbf{e}_d^{\dag}\cdot\overline{\overline{G}}(\omega_L,\mathbf{r}_{0n}^{\bot})\cdot\mathbf{e}_d],
\nonumber\\
\Gamma&=&3\gamma\lambda\sum_{n\neq 0} \mathrm{Im}[\mathbf{e}_d^{\dag}\cdot\overline{\overline{G}}(\omega_L,\mathbf{r}_{0n}^{\bot})\cdot\mathbf{e}_d]=\gamma\frac{3}{4\pi} \frac{\lambda^2}{a^2}-\gamma,
\nonumber\\
\label{D}
\end{eqnarray}
where the last equality in the expression for $\Gamma$ is valid for $a<\lambda$, $a$ being the lattice spacing of the array \cite{coop}. Using the above simplifications in Eq. (\ref{53}) and performing an inverse 2D Fourier transform back to real-space lattice representation, we obtain,
\begin{eqnarray}
\dot{\tilde{\sigma}}_{n}&=&\left[i(\delta_L-\Delta)-\frac{\gamma+\Gamma}{2}\right]\tilde{\sigma}_{n}
\nonumber\\
&+&\sum_{s=\pm}\left[e^{isq\hat{z}_n}\Omega_{ns}+\sum_k e^{isk\hat{z}_n}\delta\hat{\Omega}_{kns}(t)\right],
\label{60}
\end{eqnarray}
with the paraxial vacuum field, $\delta\hat{\Omega}_{kns}(t) = \delta\hat{\Omega}_{k_z=sk,n}(t)$.

Returning to the equation for $\hat{p}_n$ in Eq. (\ref{HL}), and considering Eqs. (\ref{47a}), (\ref{50}) and the paraxial approximation, we have
\begin{eqnarray}
\dot{\tilde{p}}_{n}&=&-m\nu_n^2\hat{z}_n+\sum_{s=\pm}\left[\hbar sq i\Omega_{ns} e^{isq\hat{z}_n}\tilde{\sigma}_n^{\dag}+\mathrm{h.c.}\right]
\nonumber\\
&+&\sum_{s=\pm}\sum_k \left[\hbar sk i e^{isk\hat{z}_n}\delta\hat{\Omega}_{kns}(t)\tilde{\sigma}_n^{\dag}+\mathrm{h.c.}\right]
\nonumber\\
&-&\sum_{m \neq n}\left[ \tilde{\sigma}_n^{\dag} A'_{nm}q(\hat{z}_n-\hat{z}_m)\tilde{\sigma}_m+\mathrm{h.c.}\right].
\label{62}
\end{eqnarray}

\subsection{The renormalized atom}
The first line of Eq. (\ref{60}) displays an atomic dipole transition that is renormalized by dipole-dipole interactions with the rest of the atoms of the array, such that the transition frequency and width are supplemented by their cooperative counterparts $\Delta$ and $\Gamma$ from Eq. (\ref{D}) \cite{coop}. Considering the assumption of non-saturated atoms, the corresponding condition for an atom $n$ then reads
\begin{equation}
P_e^n=\frac{|\Omega_n|^2}{(\delta_L-\Delta)^2+\left(\frac{\gamma+\Gamma}{2}\right)^2}\ll 1,
\label{48}
\end{equation}
where $P_e^n$ is the population of the excited state.

\section{Coarse-grained dynamics of atomic motion}
We shall now take advantage of the separation of time scales between the external and internal atomic degrees of freedom (d.o.f), allowing to move to a simpler, coarse-grained dynamical picture, wherein at any given time-bin the fast internal d.o.f reaches a steady-state that determines the evolution of the slow external d.o.f. (adiabatic elimination of the internal d.o.f). The main assumptions are therefore the conditions for timescale separation in Eqs. (\ref{sp1}) and (\ref{sp2}) below. The main results are the steady-state solution for the atomic dipole, Eq. (\ref{77}), and the resulting collective diffusion equation,  Eq. (\ref{90}). The latter describes the motion of the array atoms in the coarse-grained dynamical picture, and constitutes the main formal result of this paper, on which all subsequent analysis is based.

\subsection{Separation of time scales}
The relevant time-scale for the internal d.o.f can be read off the Eq. (\ref{60}) to be
\begin{equation}
\tau_s\sim (\gamma+\Gamma)^{-1},\Omega^{-1},(\delta_L-\Delta)^{-1}.
\label{tauc}
\end{equation}
From the equation for $\hat{p}_n$ in (\ref{62}), we can infer that the light-induced forces are of order $(1/\tau_s)\hbar q_z$. Then, for a short evolution time $T$ of the external d.o.f around some time $t_0$, we can write,
\begin{equation}
\hat{z}_n(t_0+T)\approx \hat{z}_n(t_0)+\frac{\hat{p}_n(t_0)}{m}T+\frac{\hbar q_z}{m \tau_s}T^2-\nu_n^2\hat{z}_n(t_0)T^2,
\label{zt}
\end{equation}
where the last term accounts for the force due to the trap.

\emph{Coarse-graining time.---}
In what follows, we wish to move to a coarse-grained dynamical picture with a time resolution $T$, with respect to which the internal d.o.f are fast and the external d.o.f are slow. We therefore choose $T$, such that within a time-bin $T$, the phase $q\hat{z}_n$ does not change appreciably, whereas  $\tilde{\sigma}_{n}$ does. In particular, we take the last two term in Eq. (\ref{zt}), of order $T^2$, to be much smaller than $1/q$ while $T\gg \tau_s$, which requires
\begin{equation}
\tau_s\ll T \ll \sqrt{\frac{\hbar}{E_R}\tau_s},\sqrt{\frac{\hbar}{E_R}\nu_n^{-1}}q x_{0n},
\label{T}
\end{equation}
with $x_{0n}=\sqrt{\hbar/(2m\nu_n)}$ being the zero-point motion in the traps [and using $\hat{z}_n(t_0)\sim x_{0n}$ in the last term]. This condition can be satisfied by demanding
\begin{equation}
\frac{1}{\tau_s} \gg \frac{E_R}{\hbar},\nu_n \quad \mathrm{with}\quad   E_R=\frac{\hbar^2q^2}{m}.
\label{sp1}
\end{equation}
Here $E_R$ is the recoil energy due to the laser, which yields the effective time-scale of the laser-induced evolution of the external d.o.f.  Therefore, coarse-grained dynamics with resolution $T$ satisfying (\ref{T}), assumes the existence of the \emph{separation of time scales} between the internal and external d.o.f, expressed in (\ref{sp1}), which is indeed valid in most relevant cases \cite{note2}.

\emph{Doppler effect.---}
The assumption (\ref{sp1}) allows to neglect the third and fourth terms in (\ref{zt}) for an expression for $q\hat{z}_n$. The second term gives rise to the Doppler effect, which becomes weak if we further assume that $q (\hat{p}_n/m)T\ll 1$, leading to
\begin{equation}
\frac{1}{\tau_s}\gg q\frac{\hat{p}_n}{m}.
\label{sp2}
\end{equation}
This condition entails a typically reasonable  assumption on atom velocities bounded by $\sim \lambda/\tau_s$ \cite{note3}.

\subsection{Steady-state solution for internal d.o.f}
Within any given time-bin $T$ of the coarse-grained dynamics, the fast internal d.o.f reaches a steady-state, and, subsequently can be adiabatically eliminated.
To find the steady-state solution for a given time $t$, we first formally solve Eq. (\ref{60}) from an initial time $t_0$ to a final time $t=t_0+T$. For $T$ satisfying Eq. (\ref{T}), we use $\hat{z}_n(t)$ from Eq. (\ref{zt}) excluding its $O(T^2)$ terms, finding
\begin{eqnarray}
\tilde{\sigma}_n(t)&=&e^{[i(\delta_L-\Delta)-\frac{\gamma+\Gamma}{2}](t-t_0)}\tilde{\sigma}_n(t_0)
\nonumber\\
&+& i\sum_s\Omega_{ns}\frac{e^{isq\hat{z}_n(t)}-e^{isq\hat{z}_n(t_0)}e^{[i(\delta_L-\Delta)-\frac{\gamma+\Gamma}{2}](t-t_0)}}{-i\left[\delta_L-\Delta-sq\frac{\hat{p}(t_0)}{m}\right]+\frac{\gamma+\Gamma}{2}}
\nonumber\\
&-& \sum_s \sum_{\mathbf{k}\mu} g_{\mathbf{k}\mu} e^{i\mathbf{k}_{\bot}\cdot \mathbf{r}_n^{\bot}} \left[e^{isk\hat{z}_n(t)}e^{-i(\omega_{\mathbf{k}}-\omega_L)t} \right.
\nonumber\\
&&\left. -e^{isk\hat{z}_n(t_0)}e^{-i(\omega_{\mathbf{k}}-\omega_L)t_0}e^{[i(\delta_L-\Delta)-\frac{\gamma+\Gamma}{2}](t-t_0)}\right]
\nonumber\\
&&\times \frac{1}{-i\left[\delta_L-\Delta+\omega_{\mathbf{k}}-\omega_L-sk\frac{\hat{p}(t_0)}{m}\right]+\frac{\gamma+\Gamma}{2}}\hat{a}_{\mathbf{k}\mu}.
\nonumber\\
\label{70}
\end{eqnarray}

We now define the coarse-grained observable around time $t=t_0+T$ by
\begin{equation}
\overline{\sigma}_{n}(t)\equiv\frac{1}{T}\int_{t_0}^{t_0+T}dt' \tilde{\sigma}_{n}(t').
\label{cg}
\end{equation}
Performing this coarse-graining to both sides of Eq. (\ref{70}), and neglecting terms of order $[T(\gamma+\Gamma)]^{-1}$ in accordance with (\ref{T}), we find
\begin{eqnarray}
&&\overline{\sigma}_{n}(t)=-\sum_s\Omega_{ns}\frac{e^{isq\hat{z}_n(t)}}{\delta_L-\Delta-sq\frac{\hat{p}(t)}{m}+i\frac{\gamma+\Gamma}{2}}
\nonumber\\
&&-\sum_s \sum_{\mathbf{k}\mu}\frac{i g_{\mathbf{k}\mu} e^{i\mathbf{k}_{\bot}\cdot \mathbf{r}_n^{\bot}} e^{isq\hat{z}_n(t)}e^{-i(\omega_{\mathbf{k}}-\omega_L)t}}
{\delta_L-\Delta+\omega_{\mathbf{k}}-\omega_L-sk\frac{\hat{p}(t)}{m}+i\frac{\gamma+\Gamma}{2}}\hat{a}_{\mathbf{k}\mu} \delta^T_{\omega_{\mathbf{k}},\omega_L}.
\nonumber\\
\label{74}
\end{eqnarray}
Here we used $\hat{z}_n(t')\approx \hat{z}_n(t)$ (inside the integration) and $\hat{p}_n(t_0)\approx \hat{p}_n(t)$ in accordance with the separation of time-scales from (\ref{T}) and (\ref{sp1}). The notation $\delta^T_{\omega_{\mathbf{k}},\omega_L}$ signifies a Kronecker delta of width $2\pi/T$ around $\omega_L$, which means that the vacuum modes that are included in the noise term of the second line possess frequencies contained within a bandwidth of $2\pi/T$ around $\omega_L$. The phases $e^{isk\hat{z}_n}$ of this reduced-bandwidth vacuum are correspondingly approximated as $e^{isk\hat{z}_n}\approx e^{isq\hat{z}_n}$ \cite{note4}.

Considering the condition (\ref{sp2}), we retain the dependence on $\hat{p}_n$ to lowest significant order, by taking the first term in (\ref{74}) up to linear order in $\hat{p}_n$, and to zeroth order in the weaker vacuum term, finally obtaining
\begin{eqnarray}
\overline{\sigma}_n(t)=-\sum_{s=\pm}e^{isq\hat{z}_n}\left[\frac{\Omega_{ns}+\overline{\delta\Omega}_{ns}(t)}{\delta-\Delta+i\frac{\gamma+\Gamma}{2}}
+\frac{\Omega_{ns}(sq/m)\hat{p}_n}{\left(\delta-\Delta+i\frac{\gamma+\Gamma}{2}\right)^2}\right],
\nonumber\\
\label{77}
\end{eqnarray}
with
\begin{eqnarray}
\overline{\delta\Omega}_{ns}(t)&\approx& \delta\hat{\Omega}_{ns}(t)-\frac{i}{\delta_L-\Delta+i(\gamma+\Gamma)/2}\frac{\partial}{\partial t}\delta\hat{\Omega}_{ns}(t),
\nonumber\\
\delta\hat{\Omega}_{ns}(t)&=&\sum_{\mathbf{k}_{\bot}\mu} \sum_{k_z=sk}  i g_{\mathbf{k}\mu} e^{i\mathbf{k}_{\bot}\cdot\mathbf{r}^{\bot}_n}e^{-i(\omega_{\mathbf{k}}-\omega_L)t}\hat{a}_{\mathbf{k}\mu}.
\label{80}
\end{eqnarray}
\emph{Cooperative linear-response.---} The first term of Eq. (\ref{77}) describes the linear response of a renormalized atom $n$, with cooperative shift/width $\Delta$ and $\Gamma$, to the $s=\pm$ propagating laser drive, $\Omega_{ns}$, and the corresponding vacuum field $\overline{\delta\Omega}_{ns}(t)$. The second term is the lowest-order correction due to the Doppler-effect, which is essential to obtain friction in the atomic motion (see below).
The expression for the \emph{``filtered" vacuum noise}, $\overline{\delta\Omega}_{ns}(t)$, is given by that of the usual vacuum noise $\delta\hat{\Omega}_{ns}(t)$ from Eq. (\ref{80}), but including a factor $\frac{\delta_L-\Delta+i(\gamma+\Gamma)/2}{\delta_L-\Delta+(\omega_{\mathbf{k}}-\omega_L)+i(\gamma+\Gamma)/2}\delta^T_{\omega_{\mathbf{k}},\omega_L}$, due to the atomic response. Considering the limited bandwidth of $\omega_{\mathbf{k}}$ around $\omega_L$, of order $2\pi/T\ll \gamma+\Gamma$ within the coarse-grained picture, this filter factor is approximately equal to $1$ so that $\overline{\delta\Omega}_{ns}(t)\approx \delta\hat{\Omega}_{ns}(t)$. In Eq. (\ref{80}) however, we take the lowest-order correction, of order $\frac{\omega_{\mathbf{k}}-\omega_L}{\gamma+\Gamma}$, which is formally essential to guarantee proper quantum dynamics (see subsection D below).

\subsection{Atomic motion}
Performing the coarse-graining integration (\ref{cg}) also on Eq. (\ref{62}) for $\hat{p}_n$, by again considering that $\hat{z}_n(t)$ is approximately unchanged within a time-bin $T$, we find \cite{note5}
\begin{eqnarray}
\dot{\hat{p}}_n&=&-m\nu_n^2\hat{z}_n+\hbar q\sum_{s=\pm}\left[is\overline{\sigma}_n^{\dag}e^{isq\hat{z}_n}  \left(\Omega_{ns}+\delta\hat{\Omega}_{ns}\right)+\mathrm{h.c.} \right]
\nonumber\\
&&-\sum_{m\neq n}\left[\overline{\sigma}_n^{\dag}A'_{nm}q(\hat{z}_n-\hat{z}_m)\overline{\sigma}_m+ \mathrm{h.c.}\right],
\label{81}
\end{eqnarray}
with $\delta\hat{\Omega}_{ns}$ from Eq. (\ref{80}) and $A'_{nm}$ from Eq. (\ref{47a}) (projected along the dipole orientation $\mathbf{e}_d$).

Next, we insert the solution for $\overline{\sigma}_n(t)$ from Eq. (\ref{77}) into Eq. (\ref{81}), noting the following. The second term in  $\overline{\sigma}_n(t)$, proportional to $\hat{p}_n$, leads to friction of the mechanical motion. The resulting friction coefficient includes terms due to both the classical field and the quantum vacuum field. The latter terms lead to a lower-order effect of a \emph{random} friction coefficient, which we neglect by retaining only the stronger contribution to friction arising from the classical field. By the consistency of the fluctuation-dissipation theorem, this means we should also keep only the lowest-order terms of the corresponding Langevin force. The resulting equation, Eq. (\ref{90}), is presented and discussed in the next subsection.

\subsection{Collective diffusive motion}
The assumptions and considerations mentioned above, summarized in the end of this subsection, lead to the following collective diffusion equation for the atoms,
\begin{eqnarray}
\dot{\hat{p}}_n&=&-m\nu_n^2\hat{z}_n+\bar{f}_n-\alpha_n\hat{p}_n+\hat{f}_n(t)+\sum_{m \neq n} K_{nm}(\hat{z}_n-\hat{z}_m)
\nonumber\\
&&-\sum_{m\neq n}\left(\hat{\alpha}_{nm}\hat{p}_m+\hat{p}_m\hat{\alpha}_{nm}^{\dag}\right)+\sum_{m\neq n}\hat{f}_{nm}(t),
\nonumber\\
\dot{\hat{z}}_n&=&\hat{p}_n/m.
\label{90}
\end{eqnarray}
The different terms in the first line of the equation for $\hat{p}_n$ represent the average force $\bar{f}_n$ (dipole force and dissipative force), the friction coefficient $\alpha_n$ and the corresponding Langevin force $\hat{f}_n(t)$ (due to scattering), and the mutual ``spring-constant" $K_{nm}$ (due to laser-induced dipole-dipole interactions). The two terms in the second line are usually negligible (see below) and represent the collective friction coefficient $\hat{\alpha}_{nm}$ and its corresponding collective Langevin force $\hat{f}_{nm}(t)$, both arising from cooperative scattering.

General expressions for these terms and coefficients are given in Appendix B, to leading order in $E_R/(\hbar\gamma)$ and $q\hat{z}_n$ (the latter assumed small for atoms that remain in the traps, of longitudinal extent $\sim\lambda$). For laser illumination only from the left (right-propagating, $\Omega_{ns}=\Omega_n\delta_{s+}$), they are given by
\begin{eqnarray}
\bar{f}_n&=&\hbar q|\Omega_n|^2\frac{\gamma+\Gamma}{(\delta_L-\Delta)^2+\left(\frac{\gamma+\Gamma}{2}\right)^2},
\nonumber\\
\alpha_n&=&\frac{E_R}{\hbar}|\Omega_n|^2\frac{-2(\delta_L-\Delta)(\gamma+\Gamma)}{\left[(\delta_L-\Delta)^2+\left(\frac{\gamma+\Gamma}{2}\right)^2\right]^2},
\nonumber\\
\hat{f}_n(t)&=&\hbar q\sum_{s=\pm}\left[\frac{i\Omega_n \overline{\delta\Omega}_{ns}^{\dag}(t)+is\Omega^{\ast}_n \delta\hat{\Omega}_{ns}(t)}{\delta_L-\Delta-i\frac{\gamma+\Gamma}{2}} + \mathrm{h.c.}\right],
\nonumber\\
K_{nm}&=&\frac{3}{4}\hbar q^2\gamma\left[F_{nm}\frac{\Omega_n^{\ast}\Omega_m}{(\delta_L-\Delta)^2+\left(\frac{\gamma+\Gamma}{2}\right)^2} + \mathrm{c.c.}\right].
\label{92}
\end{eqnarray}
Here, we recall $E_R=\hbar^2 q^2/m$,
and that $F_{nm}=\mathbf{e}_d^{\dag}\cdot\overline{\overline{F}}_{nm}\cdot \mathbf{e}_d$ is a dimensionless oscillatory function of the inter-atomic distances, resulting from the laser-induced dipolar interactions [see Appendix A, Eqs. (\ref{F}) and (\ref{AF})].

The correlation function of the Langevin force is found directly by using the quantum Rabi fields $\overline{\delta\Omega}_{ns}(t),\delta\hat{\Omega}_{ns}(t)$ from Eq. (\ref{80}) inside the force $\hat{f}_n(t)$ and averaging with the vacuum state, yielding,
\begin{eqnarray}
&&\langle\hat{f}_n(t)\hat{f}_m(t')\rangle=2D^{nm}_p\delta(t-t')+i2\overline{D}_p^{nm}\delta'(t-t'),
\nonumber\\
&&D^{nm}_p=(\hbar q)^2\Gamma_{nm}\frac{\Omega_n^{\ast}\Omega_m}{(\delta_L-\Delta)^2+\left(\frac{\gamma+\Gamma}{2}\right)^2},
\nonumber\\
&&\overline{D}^{nm}_p=-D_p^{nm}\frac{\delta_L-\Delta}{(\delta_L-\Delta)^2+\left(\frac{\gamma+\Gamma}{2}\right)^2},
\label{fn}
\end{eqnarray}
The second term, proportional to a derivative of a delta function, $\delta'(t-t')$, is a correction term originated in the second term of the noise $\overline{\delta\Omega}_{ns}(t)$ from Eq. (\ref{80}). If we neglect it and retain only the dominant first term, $\propto\delta(t-t')$, we get a delta-correlated Langevin force, which, although has a quantum origin (spontaneous emission), describes a purely classical Brownian motion: Since the correlation function is symmetric in $t-t'$, it corresponds to a Langevin force with vanishing commutation relations,
resulting in vanishing commutation relations for $\hat{p}_n$ and $\hat{z}_n$. The correction term $\propto\delta'(t-t')$ however, being antisymmetric, guarantees the preservation of the commutation relations, $[\hat{z}_n(t),\hat{p}_n(t)]=1$, throughout the dynamics \cite{QN}.

We note that the Langevin forces at different atomic positions $n$ are not independent, and that their correlation follows the spatial correlations contained in the vacuum field, with $\Gamma_{nm}=3\gamma\lambda\mathrm{Im}[\mathbf{e}_{d}^{\dag}\cdot G(\omega_L,\mathbf{r}_{n}^{\bot}-\mathbf{r}_{m}^{\bot})\cdot\mathbf{e}_d]$.
For an individual atom, the coefficient $D^n_p\equiv D^{nn}_p$ (noting $\Gamma_{nn}=\gamma$) is interpreted as the momentum diffusion coefficient, such that the effective temperature of the reservoir formed due to scattering is given by \cite{CCT}
\begin{equation}
T_e=\frac{D^n_p}{m\alpha_n}=-\frac{\hbar\gamma}{2}\frac{(\delta_L-\Delta)^2+\left(\frac{\gamma+\Gamma}{2}\right)^2}{(\delta_L-\Delta)(\gamma+\Gamma)},
\label{Te}
\end{equation}
which is independent of the Rabi frequency and identical for all atoms. We note that $T_e$ and $\alpha_n$ have the same sign, that should be positive if we wish to interpret the effect of scattering as damping at a rate $\alpha_n$ to a reservoir of temperature $T_e$. This requires a red cooperative detuning, $\delta_L-\Delta<0$.

The two ``collective diffusion" terms from the second line of (\ref{90}) are given in Appendix B. They are weaker than their single-atom counterparts $\alpha_n$ and $\hat{f}_n$ since they are proportional to $q(\hat{z}_n-\hat{z}_m)$, and are neglected in the following. Therefore, collective phenomena and fluctuation-dissipation phenomena are both taken to their lowest orders, i.e. via $K_{nm}$ and $\alpha_n$ (and $\hat{f}_n$), respectively.

\emph{Relation to single-atom theories.---}
We note that Eqs. (\ref{90}) and (\ref{92}) are generalizations of similar expressions found for the motion of a single atom illuminated by a laser field \cite{CCT,DAL}. Here we treat instead, the motion of an ensemble of atoms around a 2D lattice configuration. The single-body coefficients like $\bar{f}_n$, $\alpha_n$ and $D^n_p$, are similar to those obtained in the single-atom case in Ref. \cite{CCT}, except that here the ``bare",  individual-atomic resonance, with detuning $\delta_L$ and width $\gamma$, is replaced by that of the ``renormalized" atom, with $\delta_L-\Delta$ and $\gamma+\Gamma$.
The collective force term $K_{nm}$ (along with $\hat{\alpha}_{nm}$ and $\hat{f}_{nm}$) is of course totally absent in the single-atom treatments, such as that of Ref. \cite{CCT}. However, it can be directly related to the known expression for the light-induced dipole-dipole potential between pairs of atoms \cite{LIDDI}, upon the consideration of the array-renormalized atoms; see Appendix C for more details.

\emph{Summary of assumptions and approximations.---}
For completeness, we shall now briefly review the assumptions and approximations, detailed in the preceding sections, and that were used in order to arrive at Eq. (\ref{90}): (i) Atomic motion is considered here only along the longitudinal $z$ axis, assuming very deep trapping along the $xy$ plane of the 2D array (Fig. 1a,b);  (ii) We assume weak enough driving laser, such that the atoms are far from saturation, and respond linearly to light, Eq. (\ref{48}); (iii) The amplitude of the atomic motion around the 2D array geometry is assumed to be small with respect to the operating wavelength, as in Eq. (\ref{46}); (iv) The separation of time scales between internal (dipolar) and external (motional) degrees of freedom of the atoms, Eqs. (\ref{sp1}) and (\ref{sp2}), is assumed to hold. Equation (\ref{90}) is then written in  a coarse-grained dynamical picture with time resolution $T\ll E_R/\hbar,\nu_n$ (Eq. \ref{T}), fast enough to resolve the dynamics of the atomic motion; (v) For the strongest mechanical effect, paraxial illumination is assumed; (vi) We assume two-level atoms (starting from Sec. III B), with a transition dipole along the $xy$ plane. (vi) The array is assumed to be effectively infinite (practically, $\sqrt{N}\gg 1$, see Sec. V B).

\section{Collective mechanical modes}
In the previous section it was found that the motion of different atoms in the array is coupled via laser-induced dipole-dipole forces, as revealed in Eq. (\ref{90}) and depicted by Fig. 1c. We now turn to study the structure, properties and dynamics of the collective mechanical modes formed by this laser-induced coupling.

\begin{figure*}
\begin{center}
\includegraphics[scale=0.3]{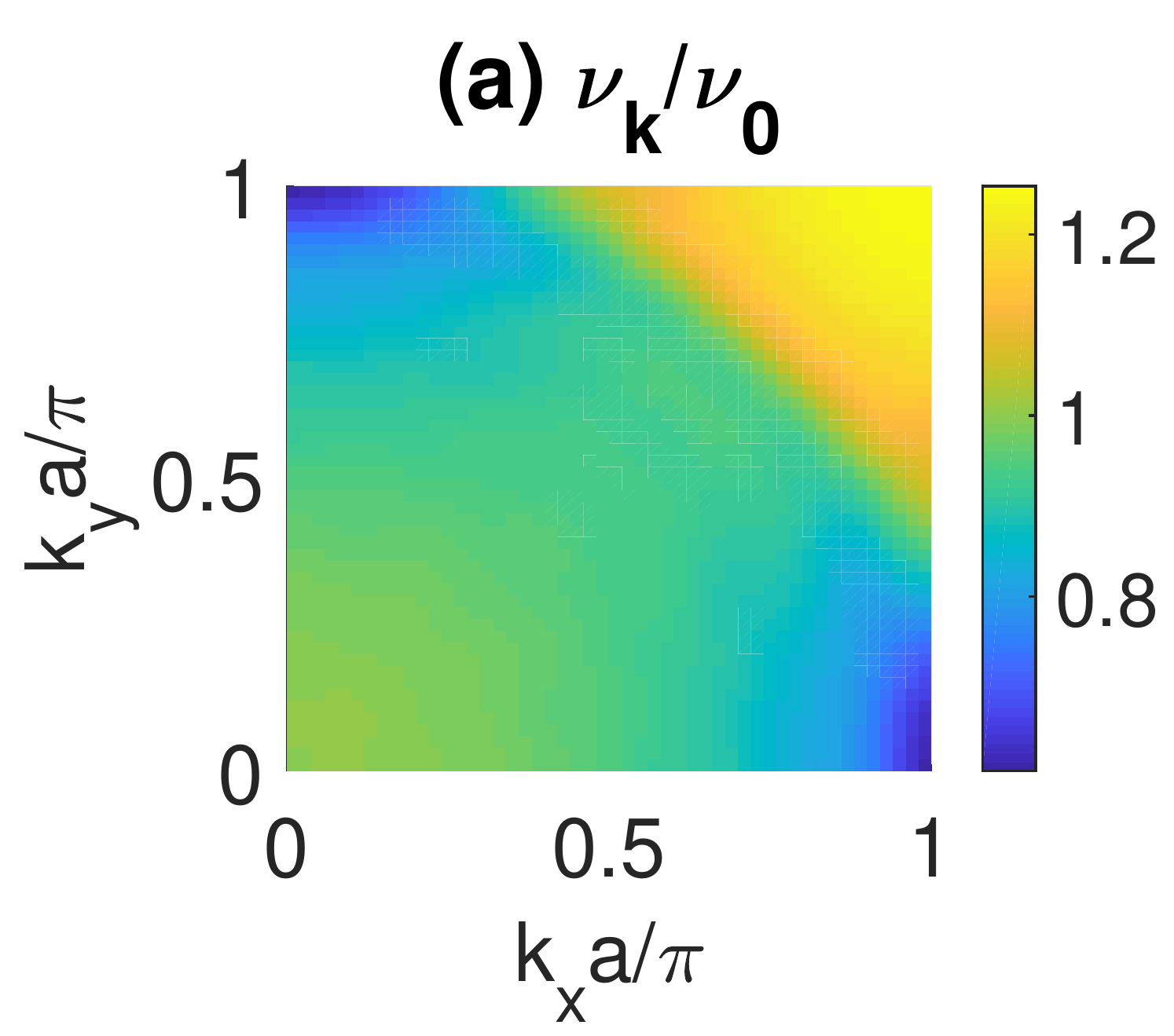}
\includegraphics[scale=0.3]{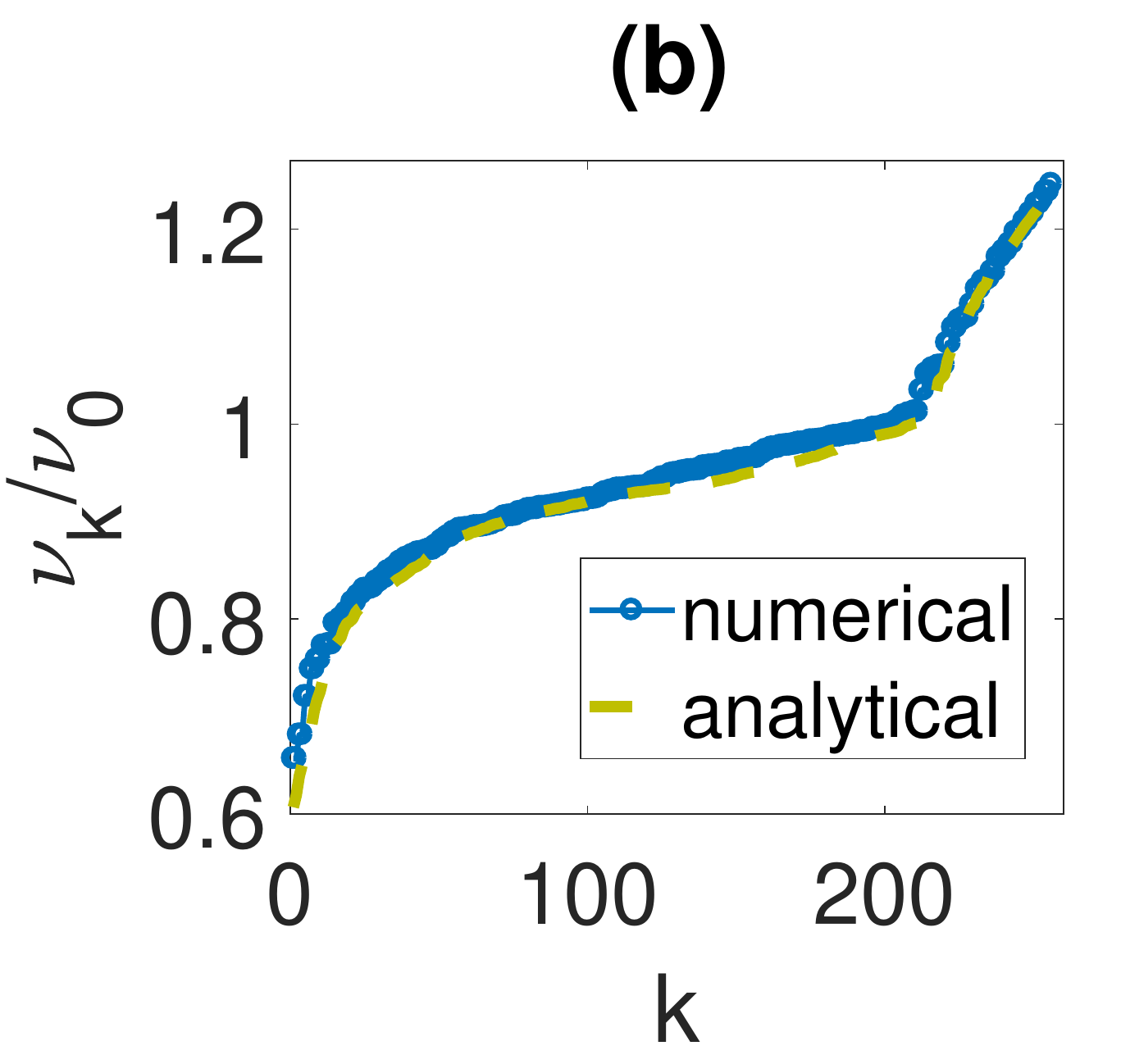}
\includegraphics[scale=0.034]{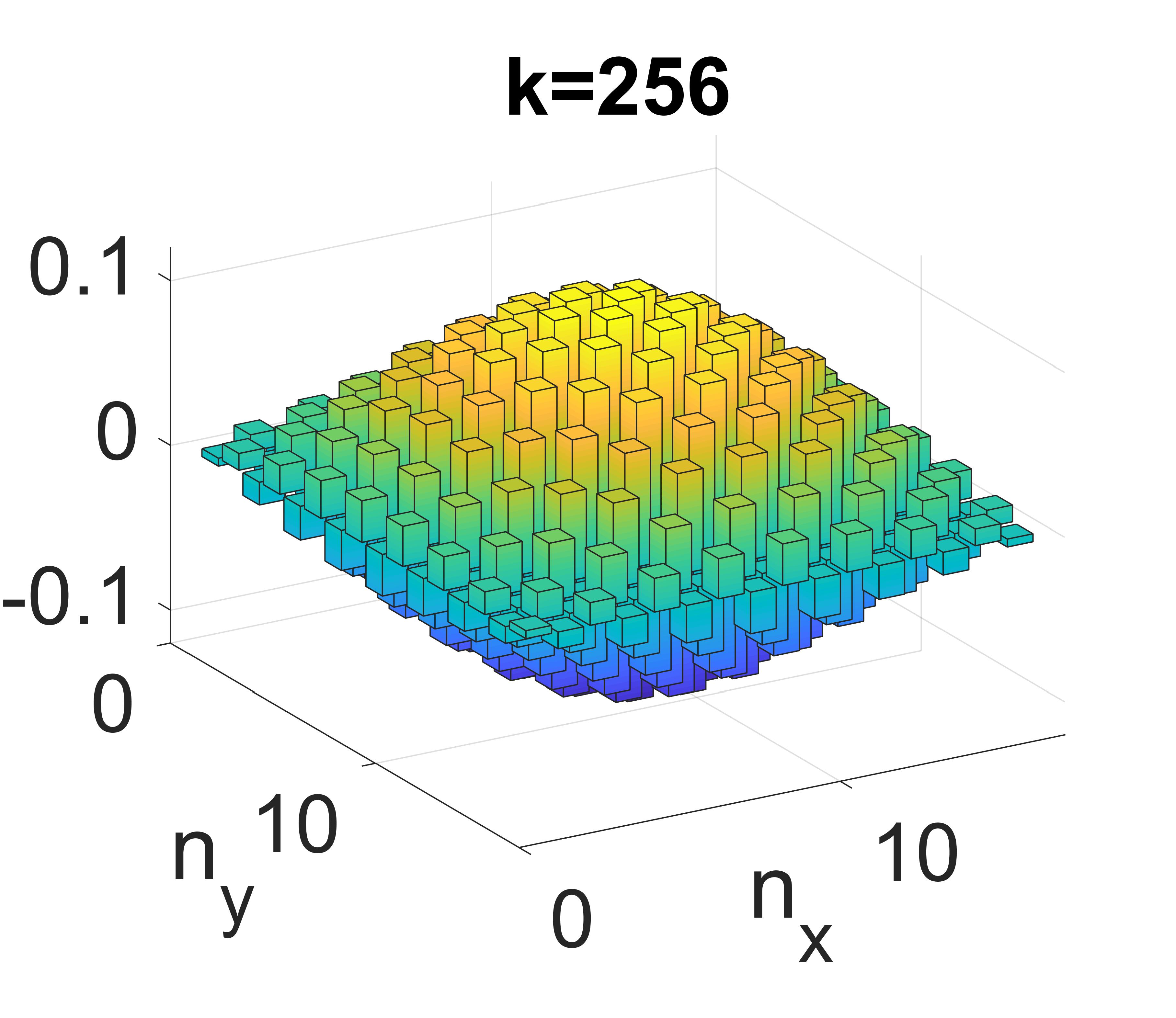}
\includegraphics[scale=0.034]{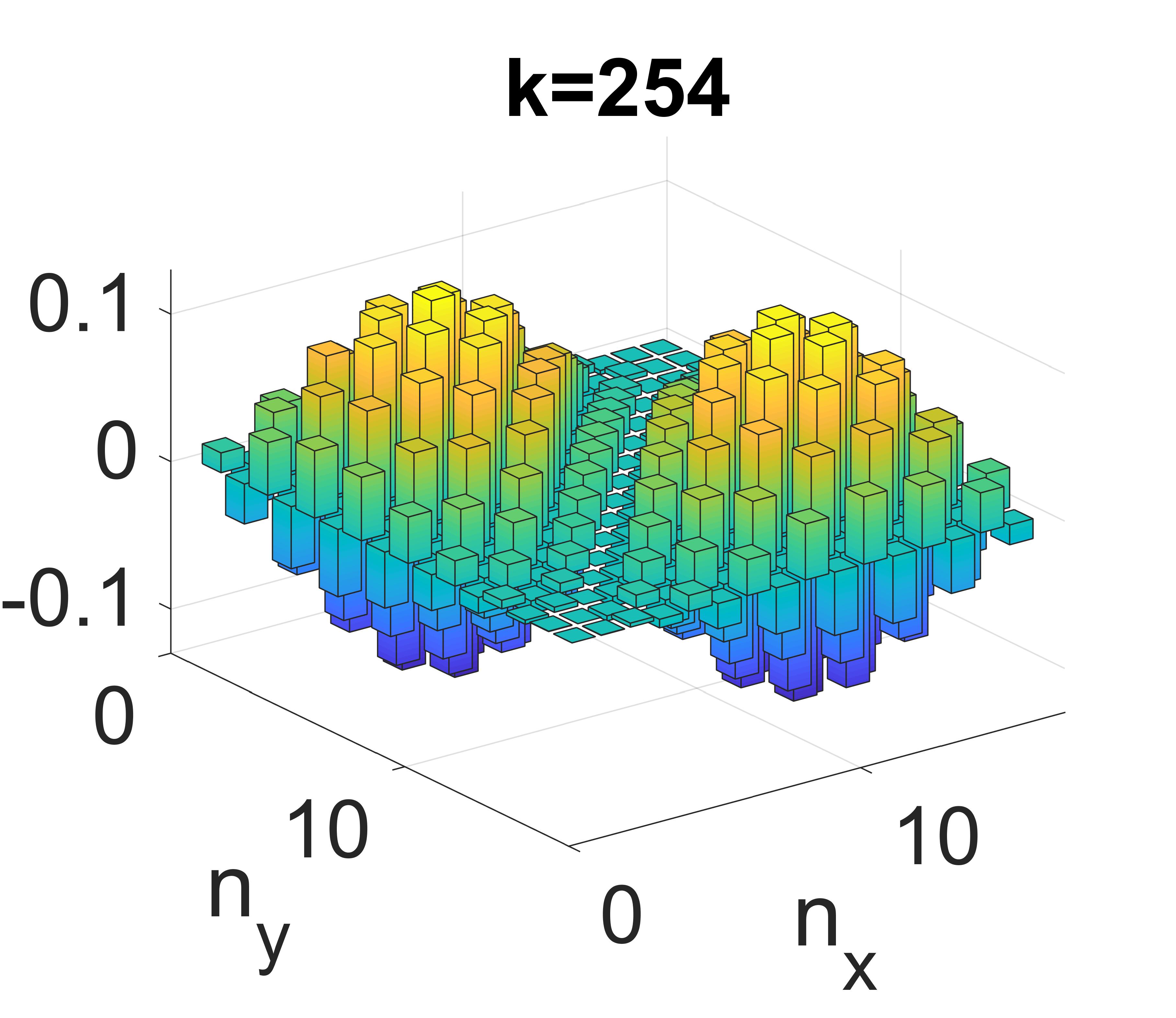}
\includegraphics[scale=0.034]{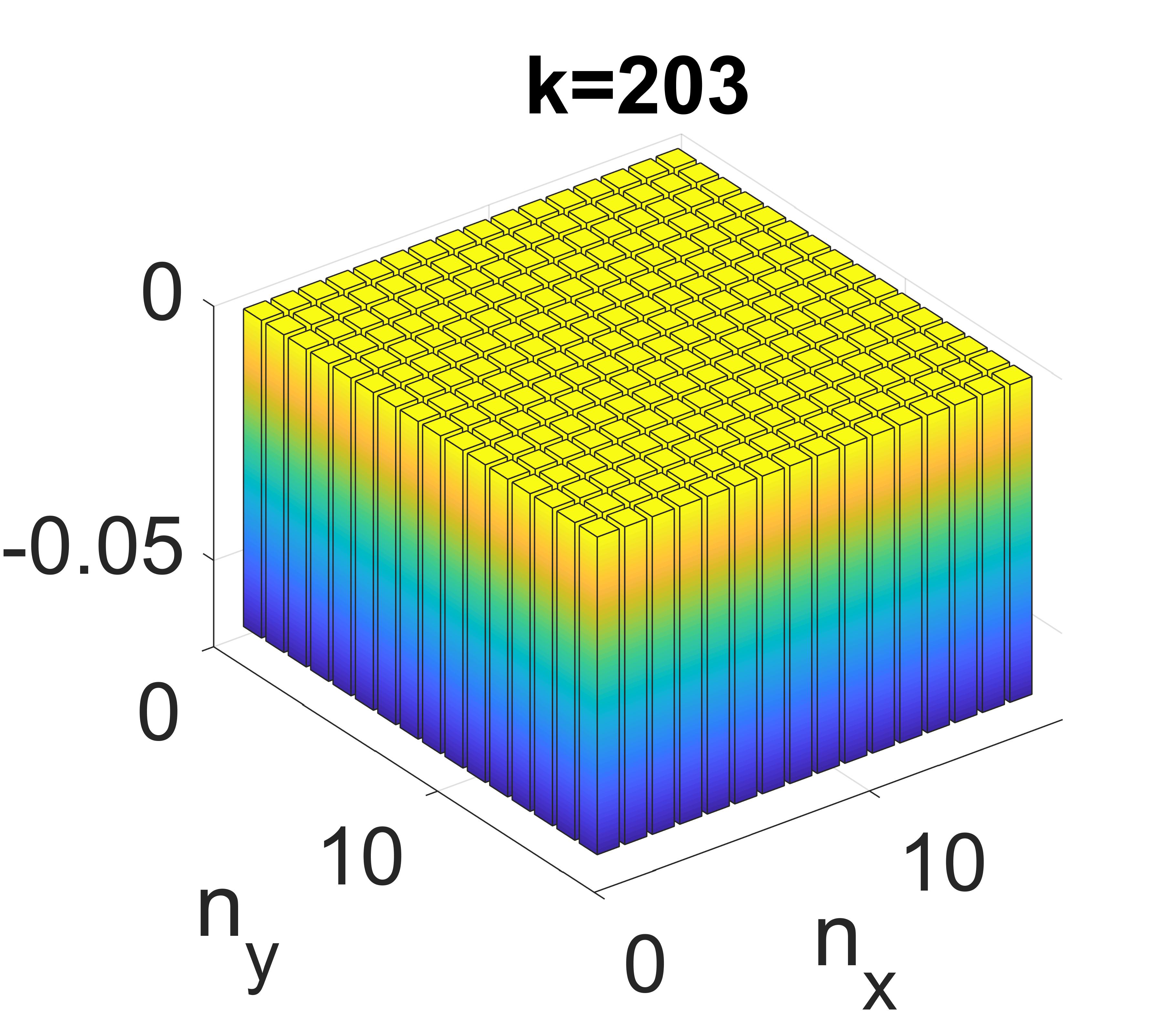}
\includegraphics[scale=0.034]{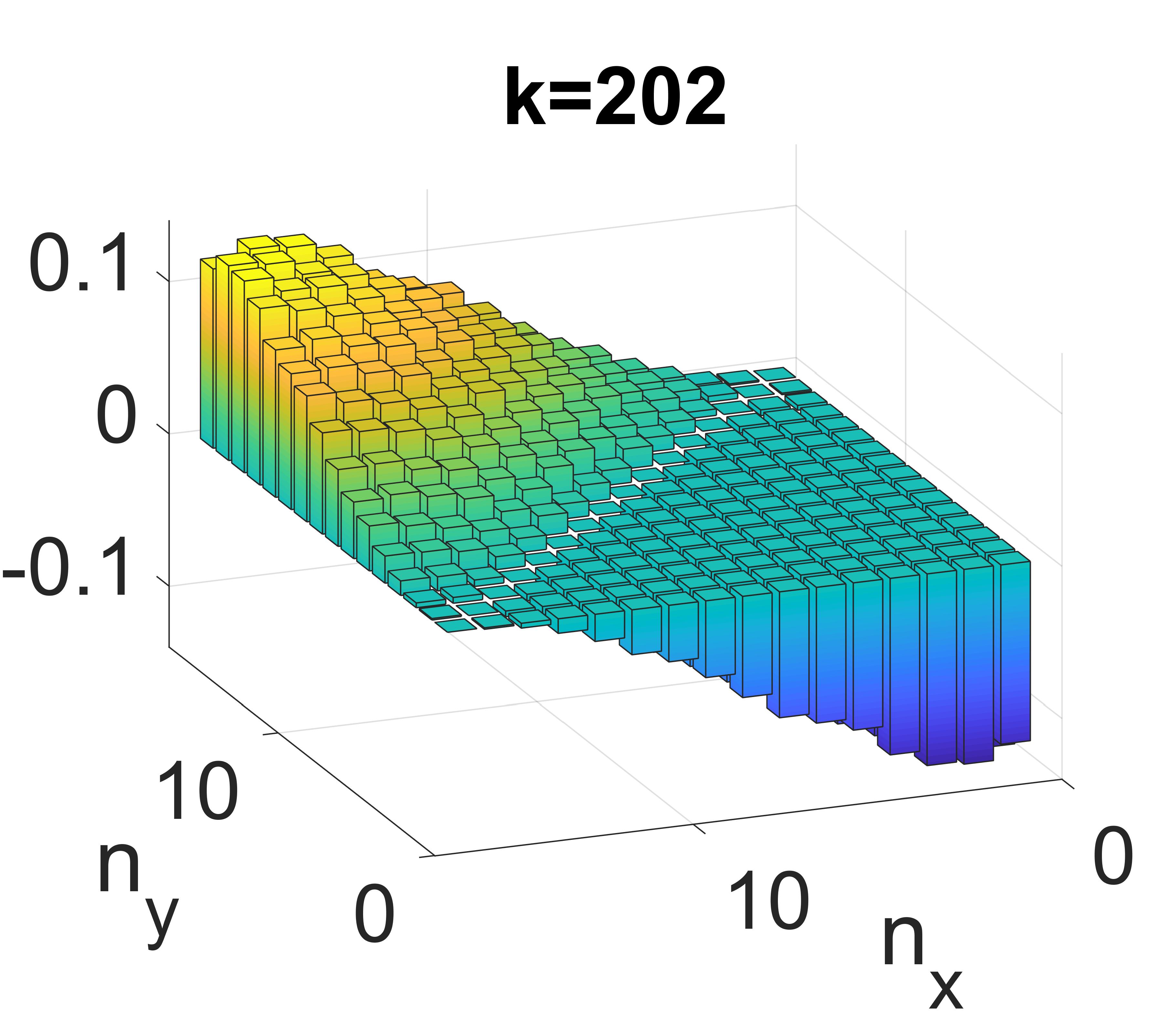}
\includegraphics[scale=0.034]{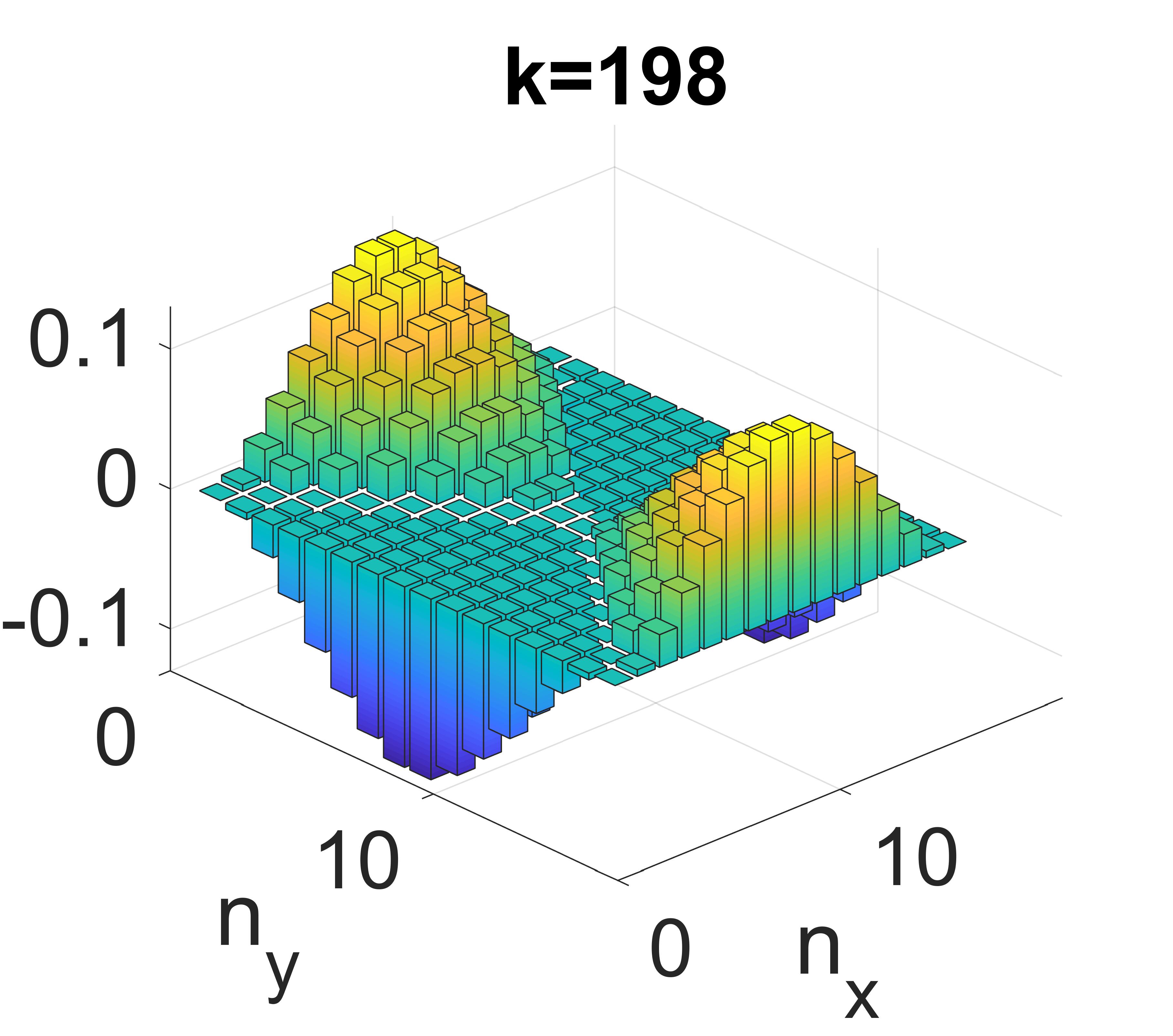}
\includegraphics[scale=0.034]{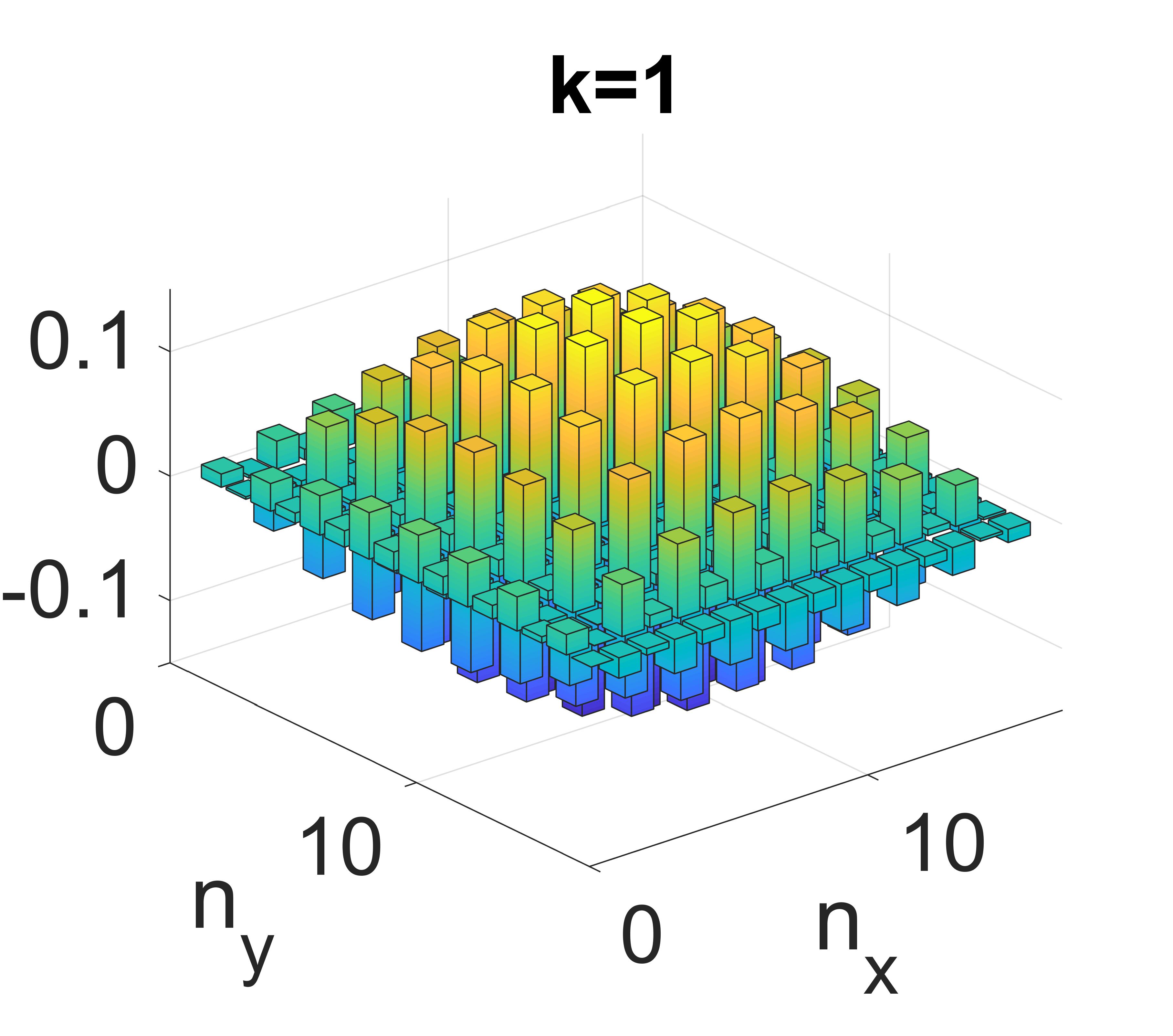}
\caption{\small{
Collective mechanical modes at uniform illumination and cooperative resonance $\delta_L=\Delta$. (a) Eigenfrequencies $\nu_{\mathbf{k}_{\bot}}$ of the normal modes $\mathbf{k}_{\bot}=(k_x,k_y)$ (2D-lattice Fourier modes) from Eq. (\ref{nuk}). (b) Agreement between the eigenfrequencies from (a) (essentially obtained analytically) and the eigenvalue spectrum $\nu_k$ of  the matrix $\overline{\overline{\nu^2}}$ from Eqs. (\ref{mat}) and (\ref{zk}), calculated numerically for a finite array of $N=16\times16$ atoms with the same array and illumination parameters as in (a). For comparison, both the eigenfrequencies from (a) and the eigenvalues $\nu_k$ are ordered in the same ascending order. The images with indices $k=256,254,203,202,198,1$ display the spatial profiles of the eigenmodes corresponding to the eigenvalues $\nu_k$ with the same $k$. They resemble 2D Fourier modes $\mathbf{k}_{\bot}$ in agreement with the analytical result from Eq. (\ref{nuk}), see text. Physical parameters used in plots: lattice spacing $a=0.5\lambda$, Rabi frequency $\Omega=0.25\gamma$ (for all atoms), $\hbar\gamma /E_R=810$ (for Rb87), and trap frequency $\nu_0\approx 10.8 E_R/\hbar$ [corresponding to trap length $l=532$nm and depth $V=200E_R$, with $\lambda=780$nm, see Eq. (\ref{nu0})].
 }} \label{fig2}
\end{center}
\end{figure*}

Rewriting Eq. (\ref{90}) in matrix form, and neglecting the weaker collective dissipative effects $\hat{f}_{nm}(t)$ and $\hat{\alpha}_{nm}$, we have
\begin{equation}
\ddot{\mathbf{z}}+\overline{\overline{\nu^2}}\mathbf{z}+\overline{\overline{\alpha}}\dot{\mathbf{z}}=\bar{\mathbf{f}}+\hat{\mathbf{f}}(t),
\label{91h}
\end{equation}
with the vector and matrix elements in the position-basis $n,n'\in [1,...,N]$ given by,
\begin{eqnarray}
\left[\mathbf{z}\right]_n&=&\hat{z}_n, \quad \left[\bar{\mathbf{f}}\right]_n=\bar{z}_n, \quad \left[\hat{\mathbf{f}}\right]_n=\hat{f}_n,
\nonumber\\
\left[\overline{\overline{\nu^2}}\right]_{nn'}&=&\left(\nu_n^2-\sum_{n''\neq n}\frac{K_{nn''}}{m}\right)\delta_{nn'}+\frac{K_{nn'}}{m},
\nonumber\\
\left[\overline{\overline{\alpha}}\right]_{nn'}&=&\alpha_n\delta_{nn'}.
\label{mat}
\end{eqnarray}
In order to analyze the mechanical normal modes of the system, it is sufficient to consider only the homogeneous, left-hand side of Eq. (\ref{91h}).
In the absence of light, $K_{nn'}=0$, the motion of different atoms is uncoupled. The ``normal modes" of motion are then the individual coordinates of the atoms $\hat{z}_n$ with ``eigenfrequencies" $\nu_n$. When the atoms are illuminated, light-induced interactions are turned on, $K_{nn'}\neq 0$, and collective mechanical modes emerge. Ignoring the friction, the resulting normal modes $\hat{z}_k$ with eigenfrequencies $\nu_k$ are found by diagonalizing the Hermitian matrix $\overline{\overline{\nu^2}}$, as
\begin{equation}
\hat{z}_k=\sum_n U_{kn}^{\ast}\hat{z}_n,   \quad  U_{kn}=\left[\mathbf{U}_k\right]_n,   \quad  \overline{\overline{\nu^2}}\mathbf{U}_{k}=\nu_k^2 \mathbf{U}_{k}.
\label{zk}
\end{equation}
As explained below, it turns out that in typical cases, the friction matrix $\overline{\overline{\alpha}}$ is approximately diagonal in the same basis, so that the eigenmodes of $\overline{\overline{\nu^2}}$ can be effectively considered as the normal modes of Eq. (\ref{91h}) [and Eq. (\ref{90})].
In the following, we analyze different types of collective normal modes $\hat{z}_k$ depending on the laser and array parameters.

\subsection{Uniform illumination}
Assuming uniform illumination $\Omega_n=\Omega$ and identical trapping frequencies $\nu_n=\nu_0$ for all atoms, the discrete translational symmetry of an effectively infinite array can be exploited. The normal modes are then 2D lattice Fourier modes, $\hat{z}_{\mathbf{k}_{\bot}}=(1/\sqrt{N})\sum_n e^{-i\mathbf{k}_{\bot}\cdot \mathbf{r}_n^{\bot}}\hat{z}_n$, i.e. $U_{\mathbf{k}_{\bot}n}=(1/\sqrt{N})e^{i\mathbf{k}_{\bot}\cdot \mathbf{r}_n^{\bot}}$, with $\mathbf{k}_{\bot}=(k_x,k_y)$ inside the first Brillouin zone, $k_{x,y}\in [-\pi/a,\pi/a]$ for a square lattice with spacing $a$. Performing the 2D Fourier transformation on Eq. (\ref{90}) [or  Eq. (\ref{91h})] we find the spectrum of the collective $\mathbf{k}_{\bot}$ modes
\begin{equation}
\nu_{\mathbf{k}_{\bot}}=\sqrt{\nu_0^2+\frac{1}{m}(K_{\mathbf{k}_{\bot}}-K_0)},
\quad K_{\mathbf{k}_{\bot}}=\sum_{n\neq 0} K_{n0}e^{-i\mathbf{k}_{\bot}\cdot \mathbf{r}_n^{\bot}},
\label{nuk}
\end{equation}
with $n=0$ denoting the central atom in the array at $\mathbf{r}_0^{\bot}=0$, $K_0=K_{\mathbf{k}_{\bot}=0}$, and where the sum in $K_{\mathbf{k}_{\bot}}$ can be obtained numerically using $K_{nm}$ from (\ref{92}).

The collective mechanical spectrum $\nu_{\mathbf{k}_{\bot}}$ is plotted in Fig. 2a at cooperative resonance $\delta_L=\Delta$, for a square array with lattice spacing $a=0.5\lambda$ and trap frequency $\nu_0\approx 10.8 E_R/\hbar$. The chosen trap frequency corresponds to realistic optical-lattice trapping with a longitudinal confinement (trap size or period along $z$) $l\approx 0.68\lambda$ ($\approx 532$nm for Rb87), trap depth $\tilde{V}=V/E_R=200$, and resulting Lamb-Dicke parameter $\eta\approx0.21$, via the relations
\begin{equation}
\nu_0=\frac{\hbar}{2m x_0^2}=\frac{1}{2\eta^2}\frac{E_R}{\hbar}, \quad \eta=q x_0=\frac{l}{\lambda}\frac{1}{(\tilde{V}/2)^{1/4}},
\label{nu0}
\end{equation}
where $x_0$ denotes the zero-point motion in the trap.

For a laser drive of $\Omega=0.25\gamma$, we observe in Fig. 2a significant deviations of  $\nu_{\mathbf{k}_{\bot}}$ from the individual-trap frequency $\nu_0$, implying the importance of collective mechanical effects. In particular, we note the high-spatial-frequency modes around $(k_x,k_y)=(\pi/a,\pi/a)$ (yellow region) where strong dipolar interactions at short range $a\sim \lambda/2$ give rise to an energy (frequency) cost of about $0.25\nu_0$ above the individual-trap frequency $\nu_0$. The modes around $(k_x,k_y)=(\pi/a,0)$ [or $(k_x,k_y)=(0,\pi/a)$] (blue regions), however, involve also longer length scales for which the oscillatory dipole-dipole kernel may change sign [c.f. Eq. (\ref{AF})], leading to a substantial \emph{decrease} in energy cost. In contrast, for the low frequency modes around  $(k_x,k_y)=(0,0)$, the atoms move in a uniform fashion, such that their mutual interaction energy vanishes, leading to an eigenfrequency identical to $\nu_0$.

\subsection{Realistic finite-size array}
In order to exploit the discrete translational symmetry of the array for the calculation of both the internal d.o.f in Eqs. (\ref{53}), (\ref{77}), and the motion in Eq. (\ref{nuk}), we had to assume that the array is infinite. Such infinite-array solutions are typically valid for atoms far enough from the edges of the array (considering the effective finite range of the interactions), whereas they fail to describe well the dynamics of atoms at the edges. Therefore, we consider two typical situations wherein our infinite-array description should be applicable: (i) The case of focused illumination, where atoms in the center are driven whereas those at the edges are effectively not excited. This situation is treated in the subsection below, and has previously shown excellent agreement with infinite-array theory for scattering calculations \cite{coop}. (ii) For a large enough array, $\sqrt{N}\gg 1$, the atoms at the edges comprise a small fraction of the total number of atoms. Then, for the description of collective modes, which by definition involve many atoms, most contribution comes from the ``bulk" atoms far from the edges, which are faithfully described by the infinite-array approximation.

When the latter criterion, of a large array $\sqrt{N}\gg 1$, is satisfied, we therefore expect the infinite-array description to hold even at uniform illumination. In order to verify this, we consider a finite array of $N=16\times16$ atoms, with the same array and trapping parameters as in Fig. 2a. The illuminating laser is taken as a Gaussian beam,
\begin{equation}
\Omega_n=\Omega e^{-|\mathbf{r}_n^{\bot}|^2/w^2},
\label{Ogaus}
\end{equation}
with a waist $w=3000\lambda$ much wider than the array and with $\Omega=0.25\gamma$, so that it reproduces the uniform illumination from Fig. 2a (at cooperative resonance $\delta_L=\Delta$). The resulting eigenfrequencies $\nu_k$ are found numerically by the matrix diagonalization from Eq. (\ref{zk}) and plotted in ascending order with mode indices $k=1,...,256$ in Fig. 2b, finding excellent agreement with the analytically-found eigenfrequency spectrum from Eq. (\ref{nuk}) (ordered in a similar fashion).

The agreement with the infinite-array calculations extends also to the spatial profiles of the corresponding eigenemodes. The highest frequency mode $k=256$ appears as the highest 2D-lattice Fourier mode $(k_x,k_y)=(\pi/a,\pi/a)$ if we ignore the edge atoms, as predicted analytically; similarly, the mode $k=254$ also exhibit high spatial frequency modulations. In contrast, the modes $k=203,202,198$, for which $\nu_k\approx \nu_0$ (see Fig. 2b), exhibit very low frequency spatial modulations, with $k=203$ corresponding to uniform motion with $\nu_k=\nu_0$, as expected. Finally, the mode $k=1$ with eigenfrequency $0.66\nu_0$ corresponds to a superposition of the lowest energy modes, $(k_x,k_y)=(\pi/a,0)$ and  $(k_x,k_y)=(\pi/a,0)$ from Fig. 2a (blue region): its spatial modulation appears as $\cos 0.5(\pi/a) (x+y)$], in agreement with $\cos(\pi/a) x + \cos(\pi/a) y$.

\begin{figure}
\begin{center}
\includegraphics[scale=0.033]{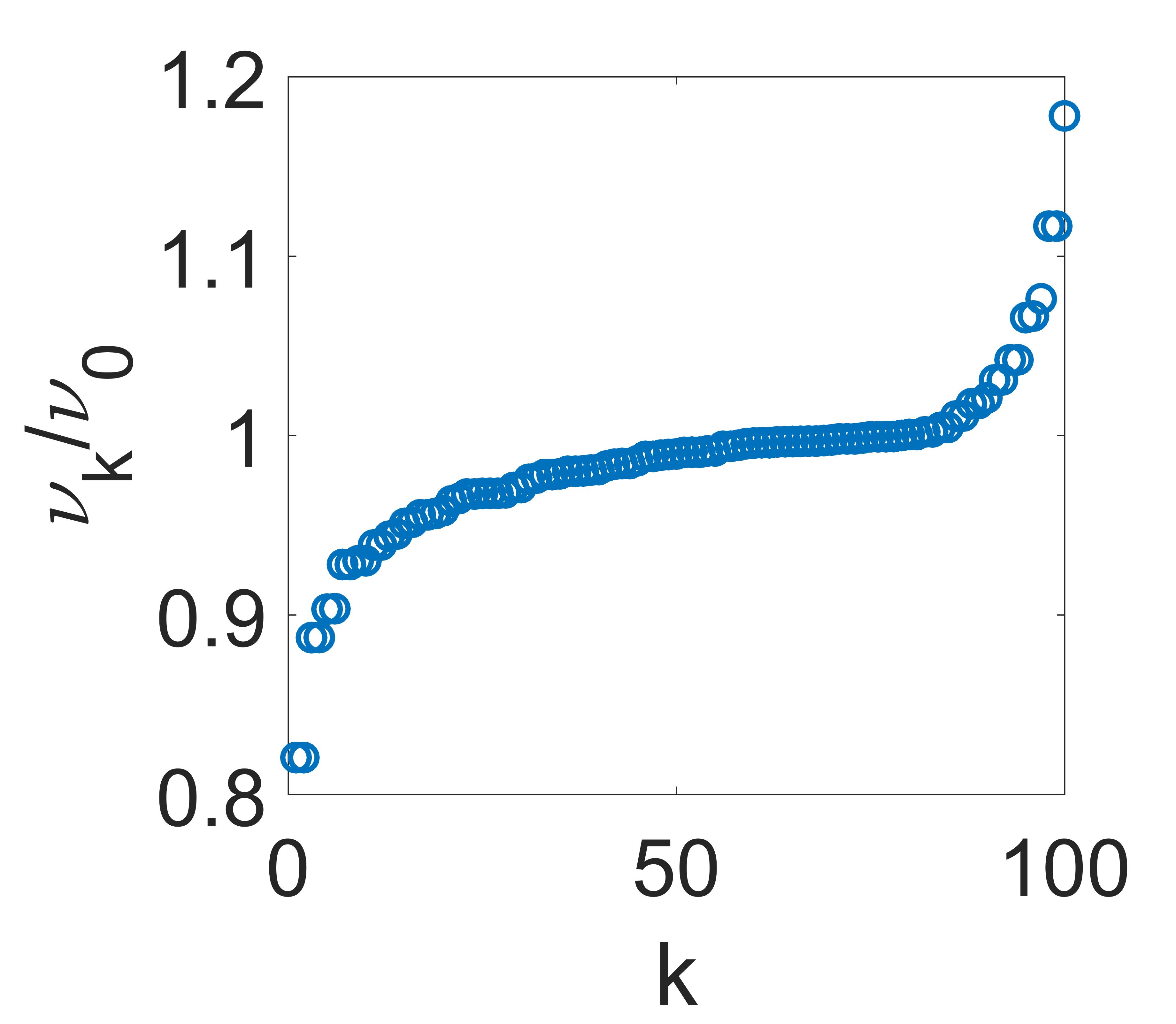}
\includegraphics[scale=0.033]{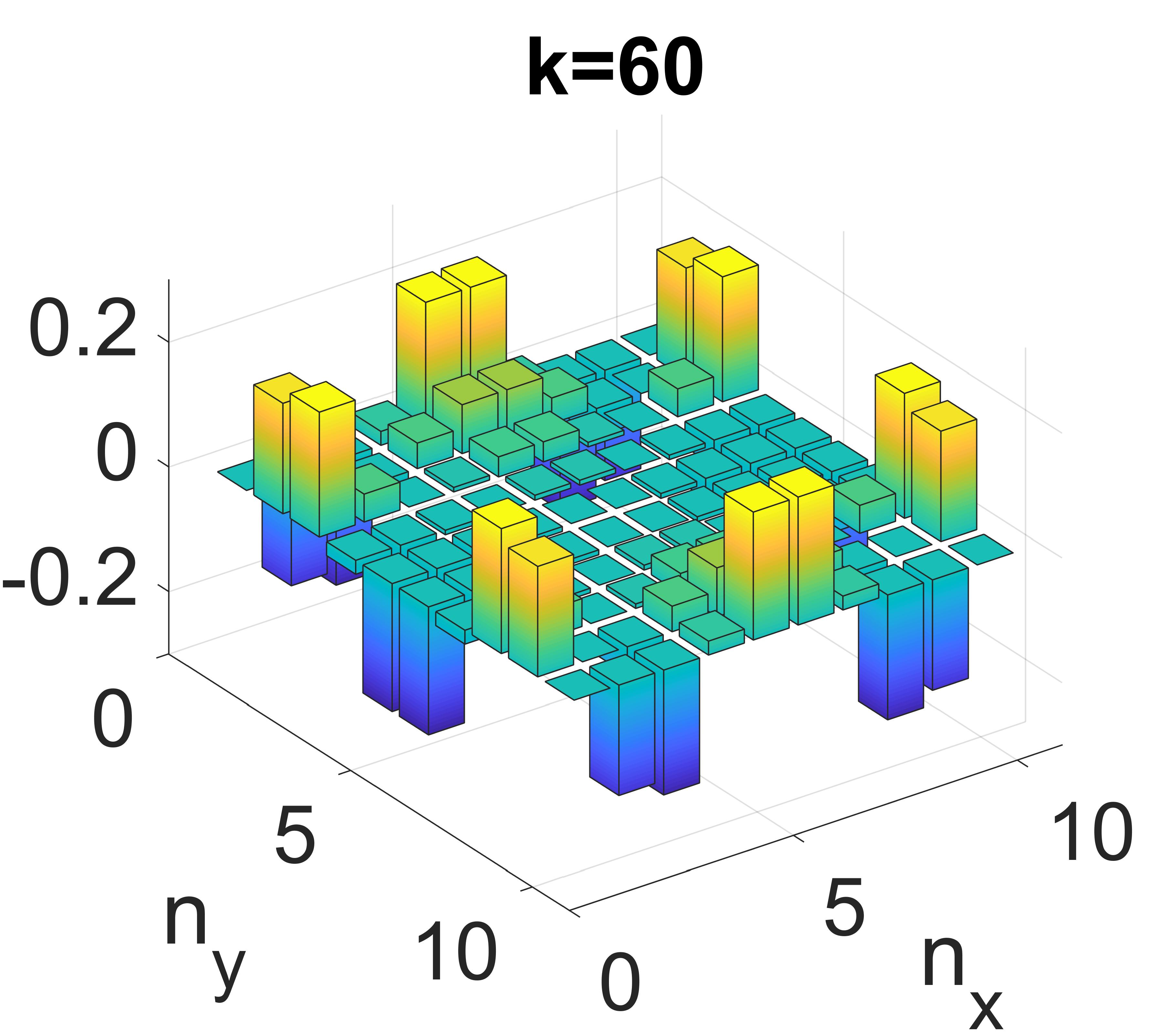}
\includegraphics[scale=0.033]{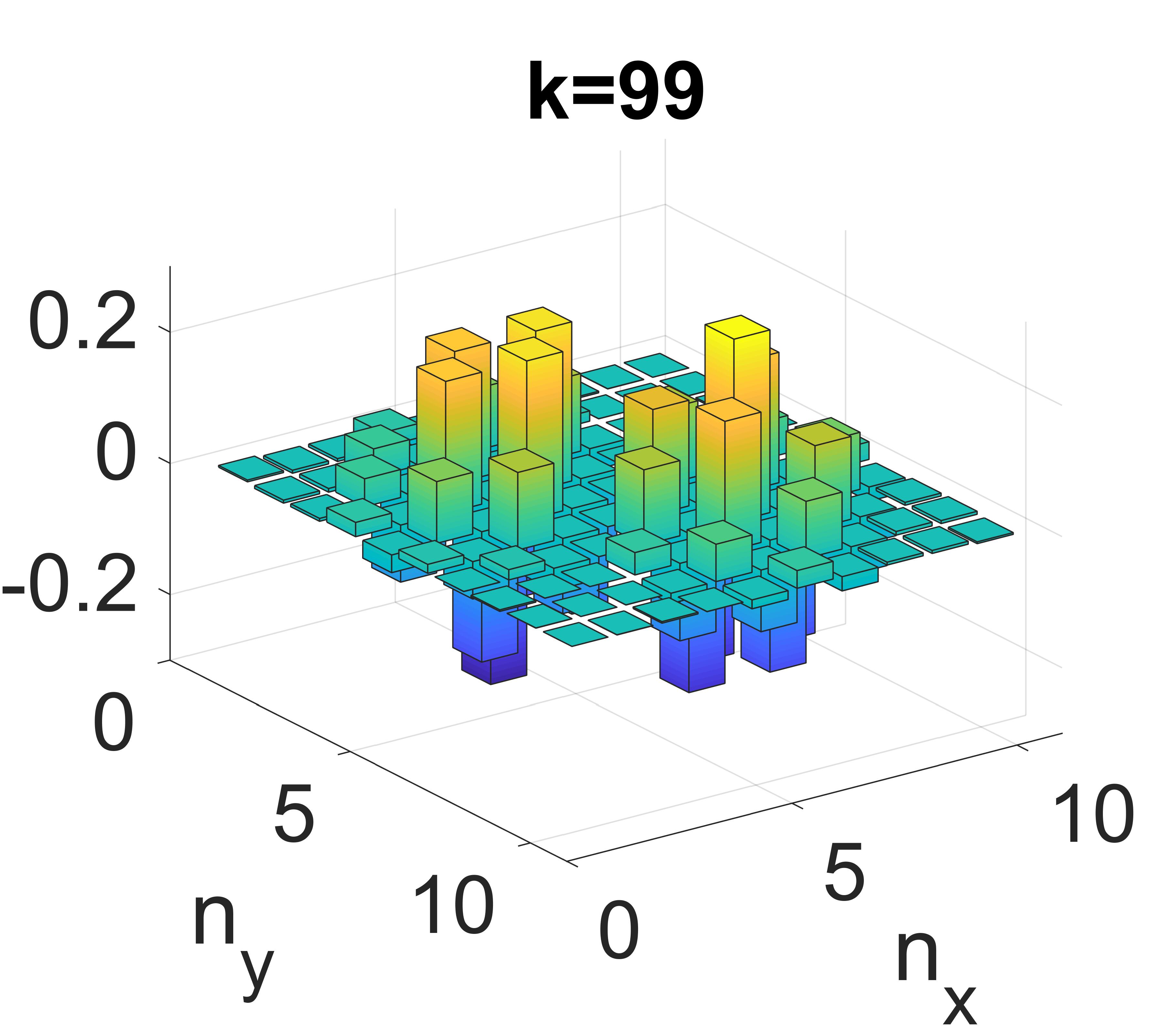}
\includegraphics[scale=0.033]{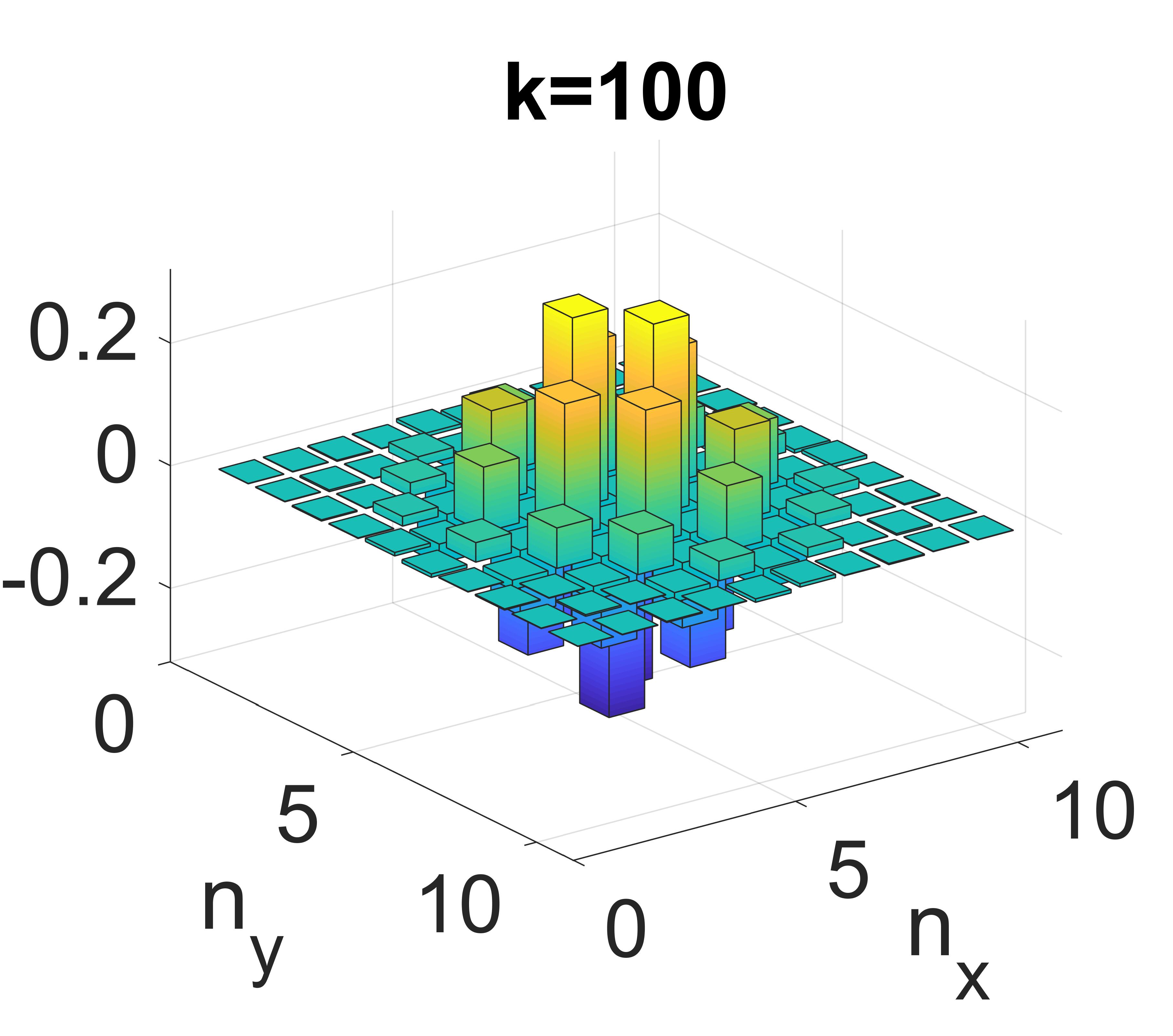}
\caption{\small{
Focused illumination and gapped modes. Here the waist of the driving Gaussian beam is smaller than the array, so that only atoms near the center are excited and experience laser-induced interactions. The collective modes formed by these several central atoms, such as the modes $k=99,100$ possess eigenfrequencies $\nu_k$ different than $\nu_0$ whereas most of the other modes (e.g. $k=60$), involving the rest of the atoms, are effectively non-interacting, with $\nu_k\approx \nu_0$. The finite array size and strong interactions at the center give rise to gapped modes such as $k=100$. Physical parameters used in this figure: $\delta_L=\Delta$, $w=2.4\lambda$, $\Omega=0.25\gamma$, $a=0.5\lambda$, $N=10\times 10$ and $\nu_0\approx 10.8 E_R/\hbar$.
 }} \label{fig3}
\end{center}
\end{figure}

\subsection{Focused illumination and gapped modes}
Consider now an illuminating Gaussian beam [Eq. (\ref{Ogaus})] with a waist $w$ that is narrower than the array size. Such a focused beam mostly excites the atoms within a radius $w$ from the center of the array (at $\mathbf{r}_{\bot}=0$) so that laser-induced interactions, and hence collective modes, are built almost only between these central atoms. As an example, we plot in Fig. 3, the eigenfrequencies $\nu_k$ and a few eigenmodes $\hat{z}_k$, for an array of $N=10\times 10$ atoms, again with $a=0.5\lambda$ and $\nu_0\approx 10.8 E_R/\hbar$, with the beam waist and central Rabi frequency being $w=2.4\lambda$ and $\Omega=0.25\gamma$. Since the array is finite and interactions are strong at the center, most of the atoms do not participate in forming collective mechanical modes. This explains the many modes around $k=50$ with an eigenfrequency close to $\nu_0$ observed in the spectrum $\nu_k$ in Fig. 3. As seen for example in the spatial profile of the mode $k=60$, these modes involve almost exclusively the atoms that are beyond the beam waist, and hence they are essentially non-interacting and equivalent to individual-atom ``modes". In contrast, the modes that involve atoms from the center, and especially those which resemble the highest spatial-frequency Fourier modes, such as $k=100$ and $k=99$ exhibit large cooperative effects. Their eigenfrequencies $\nu_k$ are different from $\nu_0$ and they may develop significant spectral gaps due to the finite size of the array (e.g. $0.062\nu_0$ between $k=100$ and $k=99$).
Therefore, a mode such as $k=100$ becomes ``rigid" in the sense that it is spectrally separated from all other modes; that is, it may be possible to selectively excite it via a proper temporal modulation to laser intensity (corresponding to a modulation of the driving radiation-pressure force).

\subsection{Unstable modes}
An interesting manifestation of the collective aspect of atomic motion is the formation of unstable mechanical modes for certain laser and array parameters. Referring back to the intuitive spring model from Fig. 1c, unstable modes are formed when the interactions $K_{nm}$ result in a strong enough overall ``repulsion" from the equilibrium position $z_n=0$, that overcomes the individual trapping potential $\nu_n$. Such a situation is enabled by the fact that the dipole-dipole interaction kernel $F_{nn'}$ from Eq. (\ref{AF}) is oscillatory and can change sign. Mathematically, this is seen by the matrix $\left[\overline{\overline{\nu^2}}\right]_{nn'}$ from Eq. (\ref{mat}): when its diagonal elements become negative, this could potentially lead to imaginary eigenvalues $\nu_k$ and hence instability. This requires the existence of positive ``spring constants" $K_{nn'}$, namely that the force between atoms $n$ and $n'$ is repulsive, so that the overall sum on atoms $n'$ is positive. As an example, by looking at Fig. 3, where no unstable modes exist, we conclude that for $a=0.5\lambda$ and the conditions considered therein, there is an overall ``attraction" and hence stability. Changing only the lattice spacing to $a=0.8\lambda$ however, is already sufficient to give rise to unstable modes: This lattice spacing is large enough to cause sign changes in the kernel $K_{nn'}$ for nearest neighbors, which give the most significant contribution to the force. This is seen in Fig. 4 for $a=0.8\lambda$, where the rest of the parameters are kept the same as in Fig. 3. The most unstable modes are those where interactions are most significant, i.e. the most spatially modulated modes. For smaller Rabi frequencies fewer modes become unstable. This behavior also exists for uniform illumination: Changing the beam waist to $w=3000\lambda$ with the same parameters as in Fig. 4, we also find that the unstable modes are the highest 2D-lattice Fourier modes.
\begin{figure}
\begin{center}
\includegraphics[scale=0.032]{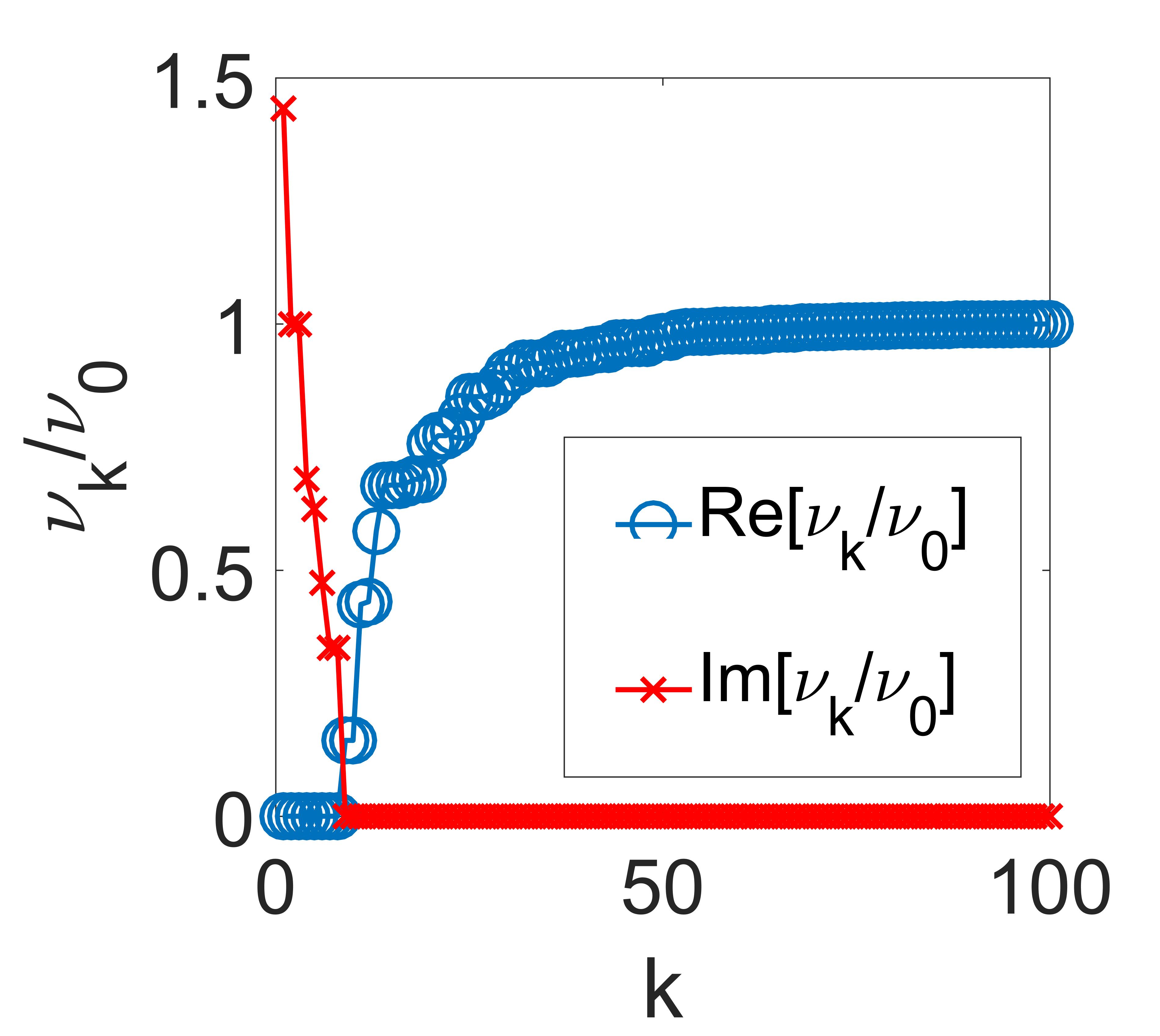}
\includegraphics[scale=0.032]{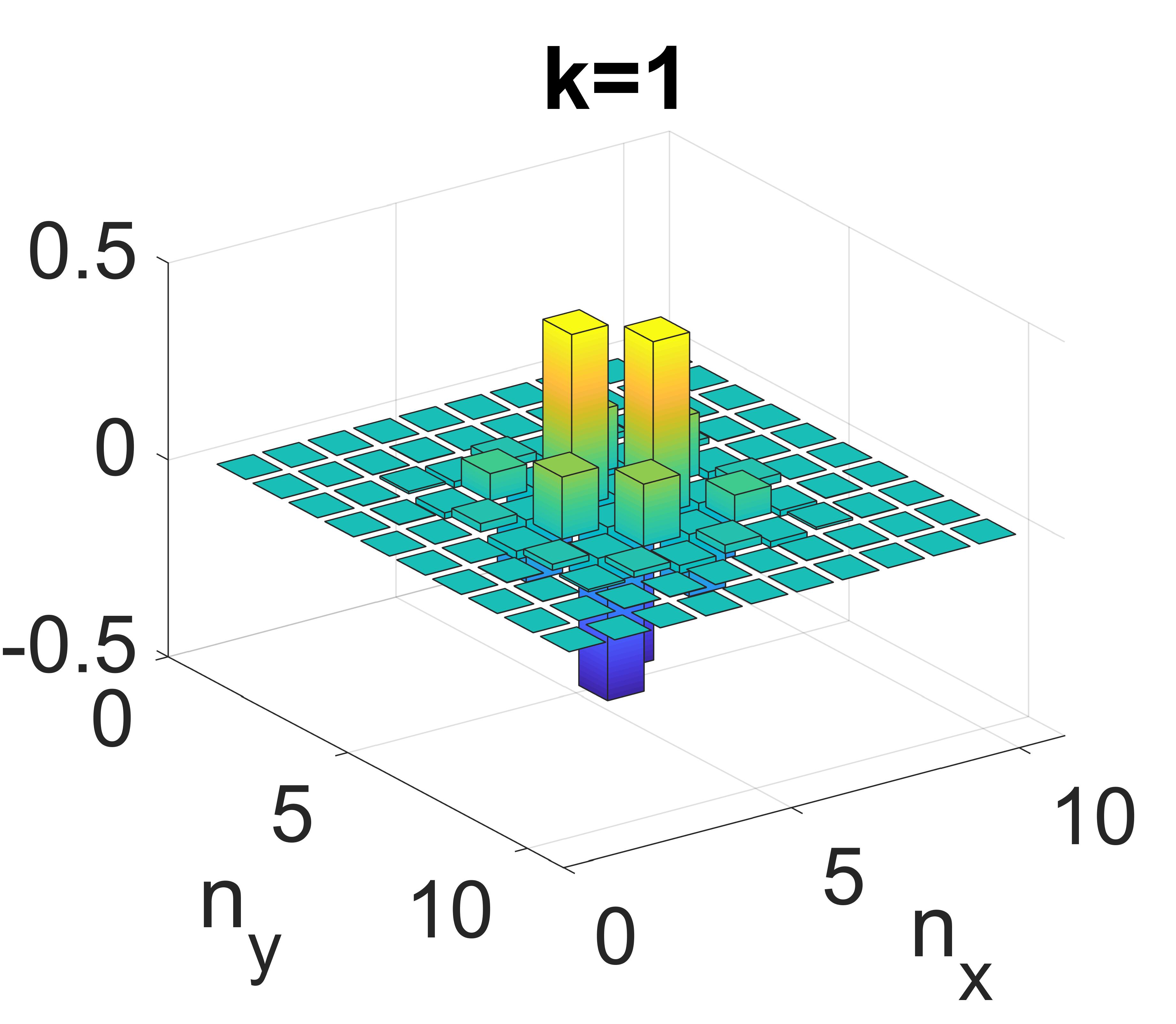}
\caption{\small{
Unstable modes: due to their oscillatory spatial dependence, laser-induced dipolar interactions can change sign and lead to unstable modes, wherein $\nu_k$ are imaginary. For the same parameters of Fig. 3, and changing only the lattice spacing to $a=0.8\lambda$, we observe the formation of collective unstable modes, primarily at high spatial modulations (e.g. the modes $k=1$).
 }} \label{fig4}
\end{center}
\end{figure}

\subsection{Dynamics of the collective modes}
Returning to the dynamical equation (\ref{91h}), we note that the role of the driving laser is two fold: First, it leads to the formation of the collective mechanical modes, i.e. the eigenmodes of the homogenous left-hand side of the equation, via laser-induced interactions; and second, it drives these same modes, via the average radiation pressure and Langevin noise at its right-hand side. Writing Eq. (\ref{91h}) in the eigenmode $k$ basis of $\overline{\overline{\nu^2}}$ via the transformation from Eq. (\ref{zk}), we have
\begin{equation}
\dot{\hat{p}}_k=-m\nu_0^2\hat{z}_k+\bar{f}_k-\alpha_k\hat{p}_k+\hat{f}_k(t),
\quad
\dot{\hat{z}}_k=\hat{p}_k/m.
\label{EOMk}
\end{equation}
Here we neglected the off-diagonal friction $\alpha_{kk'}\approx \alpha_{k}\delta_{kk'}$. We verify numerically for a variety of array and laser parameters that these off-diagonal elements are indeed substantially smaller than the diagonal ones $\alpha_k$, which are in turn much smaller than the eigenfrequencies $\nu_k$ close to cooperative resonance. As noted above, it is the spatial profile of the laser that both determines the modes $k$, and also the strength of their driving, via their spatial overlap with the laser forcing, $\bar{f}_k$.

The resulting collective diffusive dynamics is found by moving to shifted collective coordinates $\hat{z}'_k=\hat{z}_k-\bar{f}_k/(m \nu_k^2)$, and solving for the set of two linear differential equations (\ref{EOMk}), finding,
\begin{eqnarray}
\hat{z}_k(t)&=&\bar{z}_k+e^{-(\alpha_k/2)t}\sin(\tilde{\nu}_k t)\frac{\hat{p}_k(0)}{m\tilde{\nu}_k}
\nonumber\\
&+&e^{-(\alpha_k/2)t}\left[\cos(\tilde{\nu}_k t)+\frac{\alpha_k}{2\tilde{\nu}_k}\sin(\tilde{\nu}_k t)\right]\left(\hat{z}_k(0)-\bar{z}_k\right)
\nonumber\\
&+&\int_0^tdt' e^{-(\alpha_k/2)(t-t')}\sin[\tilde{\nu}_k (t-t')]\frac{\hat{f}_k(t')}{m\tilde{\nu}_k},
\label{zsol}
\end{eqnarray}
with $\tilde{\nu}_k=\sqrt{\nu_k^2-(\alpha_k/2)^2}$ and $\bar{z}_k=\bar{f}_k/(m \nu_k^2)$. Together with the transformation $U_{kn}$ to position-basis, Eq. (\ref{zk}), the initial zero-point motion in the traps $\langle \hat{z}_n(0)^2\rangle=x_{0n}^2$, $\langle \hat{p}_n(0)^2\rangle=m^2\nu_n^2x_{0n}^2$ ($x_{0n}$ being the zero-point motion in the trap $n$), and the Langevin-force correlation function, Eq. (\ref{fn}), the statistics of the motion can be fully determined. This is demonstrated in the next section for the case of the estimation of heating of the irradiated atoms.

\section{Example: heating of the atoms}
As mentioned in the Introduction, ordered atomic arrays were recently considered as a promising platform for quantum optics, relying on their cooperative dipolar response \cite{coop,janos,ADM2,janos2,ABA,CHA,ANA,HEN,ADM}. Specific applications of 2D arrays, such as nearly-perfect reflectors  \cite{coop,ADM}, or efficient quantum memories \cite{CHA}, involve normal incident light near cooperative resonance, $\delta_L\approx\Delta$, where the mechanical effect of light is maximized.
In particular, resonant light may heat the atoms, so that they could escape the traps, effectively ``melting" down the 2D array. In the following, we apply the formalism developed above, in order to estimate the heating and resulting motion of the atoms. Considering continuous laser drive, and depending on the detuning from cooperative resonances and the trap depth, we discuss the conditions for the thermalization of the atoms inside the traps. For cases where the atoms escape the traps before thermalization, we estimate the escape time, at which the atom array effectively melts down, and the total number of photons reflected by the array until this time.

For simplicity of the presentation, we neglect in the following the effect of the dipole-dipole forces, $K_{nm}$, on the calculation of averages and variances of the motion, $\hat{z}_n$. This should be valid for sufficiently weak laser drive (where $K_{nm}<m\nu_n^2$), and should suffice for a rough estimation of the heating. While this means that the formation of the mechanical collective modes becomes irrelevant for the analysis below, collective effects still take significant part due the renormalized atomic response.

Setting $K_{nm}= 0$ in Eq. (\ref{90}) (and ignoring its negligible second line), the solution for $\hat{z}_n(t)$ is given by Eq. (\ref{zsol}) with the normal mode index $k$ replaced by the atom index $n$. The mean and variance of $\hat{z}_n$ are then found by performing averages with the initial ground-state of the trap and the vacuum for the field, using $\langle\hat{z}_n(0)^2\rangle=x_{0n}^2$, $\langle \hat{p}_n(0)^2\rangle=m^2\nu_n^2x_{0n}^2$ and Eq. (\ref{fn}) [neglecting the small $\propto\delta'(t-t')$ correction],
\begin{eqnarray}
&&\langle\hat{z}_n(t)\rangle=\frac{\bar{f}_n}{m\nu^2_n}\left[1-e^{-(\alpha_n/2)t}\left(\cos\tilde{\nu}_n t+\frac{\alpha_n}{2\tilde{\nu}_n}\sin\tilde{\nu}_n t\right)\right],
\nonumber\\
&&\mathrm{Var}[\hat{z}_n](t)=e^{-\alpha_n t}\left(\cos\tilde{\nu}_n t+\frac{\alpha_n}{2\tilde{\nu}_n}\sin\tilde{\nu}_n t\right)^2 x_{0n}^2
\nonumber\\
&&+e^{-\alpha_n t}\left(\frac{\nu_n}{\tilde{\nu}_n}\sin\tilde{\nu}_n t\right)^2x_{0n}^2
+\frac{D_p^n}{m^2\tilde{\nu}_n^2\alpha_n}\left(1-e^{-\alpha_n t}\right)
\nonumber\\
&&-\frac{D_p^n\alpha_n}{4m^2\tilde{\nu}_n^2\nu_n^2}\left[1-e^{-\alpha_n t}\left(\cos(2\tilde{\nu}_n t)-2\frac{\tilde{\nu_n}}{\alpha_n}\sin(2\tilde{\nu}_n t)\right)\right].
\nonumber\\
\label{B25}
\end{eqnarray}
Below, we discuss two distinct scenarios, wherein the atoms either equilibrate within the traps, or escape the traps before equilibration (the effectively frictionless case).

\subsection{Thermalization case}
The atomic motion reaches a steady-state provided that the friction $\alpha_n$ is positive. This requires red-detuned light with respect to the cooperative resonance, $\delta_L<\Delta$ [c.f. Eq. (\ref{92})].
At times much longer than the equilibration time, $t\gg 1/\alpha_n$, the dynamics reach steady-state and the average motion from Eq. (\ref{B25}) become a static shift,
\begin{eqnarray}
\langle\hat{z}_n\rangle(t)&=&\bar{z}_n=\frac{\bar{f}_n}{m\nu^2_n}=\lambda \frac{3}{2\pi^2}\left(\frac{\lambda}{a}\right)^2\frac{\hbar\gamma}{E_R}\eta^4 P_e^n.
\label{av}
\end{eqnarray}
In the last equality, we assumed identical traps $\nu_0$ with corresponding Lamb-Dicke parameter $\eta$ [Eq. (\ref{nu0})], and used the expressions for $\bar{f}_n$ and $\Gamma$ from Eqs. (\ref{92}) and (\ref{D}), respectively. For tight enough trapping, the Lamb-Dicke parameter $\eta$ is small, which can make the shift $\bar{z}_n$ much smaller than a wavelength, so that the average motion is contained inside the trap (of length $l\sim \lambda$). For example, for the array parameters considered above, with $a=0.5\lambda$, $\hbar\gamma/E_R\approx810$ (e.g. for Rb87) and $\eta=0.21$ (corresponding to potential depth $V=200E_R$ and trap size/length of $l=0.68\lambda$), we obtain $\bar{z}_n/\lambda\approx 1.07 P_e^n$, which is very small for non-saturated atoms, $P_e^n\ll 1$, as assumed above.

For the variance at equilibrium, $t\gg 1/\alpha_n$, we obtain form (\ref{B25}),
\begin{eqnarray}
\mathrm{Var}[\hat{z}_n](t)&=&\frac{D_p^n}{m^2\tilde{\nu}_n^2\alpha_n}+\frac{D_p^n}{2m^2\tilde{\nu}_n \nu_n^2}\sin(2\tilde{\nu}_n t)-\frac{D_p^n\alpha_n}{4m^2\tilde{\nu}_n^2\nu_n^2}.
\nonumber\\
\label{var1}
\end{eqnarray}
Considering the very reasonable assumption $\nu_n\gg \alpha_n$, valid for non-saturated atoms in the Lamb-Dicke regime [c.f. Eqs. (\ref{92}) and (\ref{nu0})], we obtain $\mathrm{Var}[\hat{z}_n]\approx D_p^n/(m^2\nu_n^2\alpha_n)$, which, upon comparison to Eq. (\ref{Te}), gives
\begin{equation}
\frac{1}{2}m \nu_n^2\mathrm{Var}[\hat{z}_n]=\frac{1}{2}T_e.
\label{eqp}
\end{equation}
This equipartition relation shows that the steady-state is a thermal equilibrium with the expected temperature $T_e$ of the effective scattering bath. This result is valid provided that the harmonic approximation for the potential holds, restricting the temperature to be sufficiently smaller than the potential depth $V$. In contrast, for $T_e\gtrsim V$, the atoms escape the trap before they reach equilibrium, as further discussed in the next subsection.

As explained above, for the atoms to remain in the traps and equilibrate at a finite temperature $T_e$, the laser has to be off cooperative resonance, $\delta_L-\Delta<0$. Larger detunings $|\delta_L-\Delta|$ entail smaller temperatures, which are useful for satisfying $T_e < V$. On the other hand, interesting effects, such as strong reflectivity $r=-\frac{i(\gamma+\Gamma)/2}{i(\gamma+\Gamma)/2+\delta_L-\Delta}$ \cite{coop}, occur near cooperative resonance. One compromise could be to consider cooperative detunings close enough to resonance, $|\delta_L-\Delta|\ll \gamma+\Gamma$. For example, considering the $a=0.5\lambda$ array from Fig. 2, with $\delta_L-\Delta=-(\gamma+\Gamma)/4$, we find $|r|^2=0.8$ and $T_e= 506 E_R$. Increasing $V$ to exceed $T_e$ (from $200E_R$ in Fig. 2), does not change the spatial profile of the collective modes or the shape of their spectrum $\nu_k$  from Fig. 2. However, it does reduce the contrast of this spectrum: e.g. for $V/E_R=1000$ the maximal (minimal) $\nu_k/\nu_0$ becomes $1.06$ ($0.93$). This is since the ``spring-constants" ratio, $K_{nm}/(m\nu_0^2)$, decreases with $V$ [c.f. Eqs. (\ref{92}) and (\ref{nu0})].

\subsection{The effectively frictionless case}
Consider now the case where the friction is negligible with respect to all other time-scales, namely, that $\alpha_n\ll \nu_n, 1/t_e$ with $t_e$ being the duration of the experiment. This situation becomes exact at the cooperative resonance $\delta_L=\Delta$, where the friction is identically zero. In practice, however, this effectively frictionless case describes the situation where the atoms escape the traps \emph{before} thermalization, which is expected to occur for intermediate depth of the trapping potential, $T_e\gtrsim V$ (this is in contrast with an escape process activated by temperature, which occurs \emph{after} thermalization \cite{KRA}).
When friction is negligible, $\alpha_n\rightarrow 0$, the effective temperature of the ``scattering bath" is infinite, $T_e\rightarrow \infty$, c.f. Eq. (\ref{Te}), so that thermal equilibrium is never reached and the atoms are constantly heated.

Taking $\alpha_n= 0$ and hence $\tilde{\nu}_n=\nu_n$ in Eq. (\ref{B25}), the average motion becomes oscillatory, $\langle\hat{z}_n(t)\rangle=\bar{z}_n\left[1-\cos\nu_nt\right]$, with the peak-to-peak amplitude $\bar{z}_n$ from Eq. (\ref{av}), typically contained inside the trap.

Turning to the variance, we obtain from Eq. (\ref{B25}) in this case (with identical traps $\nu_n=\nu_0$)
\begin{equation}
\mathrm{Var}[\hat{z}_n](t)=x_{0}^2+D_z^n t+\frac{D_z^n}{\nu_n}\sin(2\nu_0 t)\approx D_z^n t.
\label{zva}
\end{equation}
Here $D_z^z=D_p^n/(m^2\nu_0^2)$ is identified as an effective diffusion constant for the atom $n$. The last approximate equality is valid for times longer than the trap oscillation, $t\gg 1/\nu_0$, and long enough for the variance to grow larger than the initial zero-point motion $x_0^2$.
We can define the \emph{escape time} $\tau^n_{\mathrm{esc}}$ to be that where the position fluctuations $\sqrt{\mathrm{Var}[\hat{z}_n](t)}$ reach $\lambda$,
\begin{equation}
\tau^n_{\mathrm{esc}}=\frac{\lambda^2}{D_z^n}=\gamma^{-1}\frac{\pi^2}{\eta^2 P_e^n}.
\label{tesc}
\end{equation}
Assuming a trap size to be on a wavelength scale, the escape time then provides a time scale for the ``melting" of the atom array, when the atoms escape the traps, and the mirror effect does not work anymore. It also gives a time scale at which our small-amplitude assumption, Eq. (\ref{46}), breaks down.

For an experiment where the reflection of light for the atom mirror is detected, it is therefore possible to estimate the total number of reflected photons as follows. Assuming perfect reflection (i.e. at cooperative resonance $\delta_L=\Delta$), the number of reflected photons is equal to the number of photons that hit the array before it melts. This number can be estimated by
considering the power of the incident Gaussian beam, $W=(\pi/2)w^2c\varepsilon_0 |E_0|^2$  ($E_0=\hbar\Omega/d$ being the field at the center), times the escape time $\tau^n_{\mathrm{esc}}$ (with $\Omega_n=\Omega$ for an atom at the center), and divided by the energy of an incident photon, $\hbar\omega_L=\hbar qc$, yielding
\begin{equation}
n_{\mathrm{esc}}=\frac{W\tau^n_{\mathrm{esc}}}{\hbar \omega_L}= \frac{3\pi^2}{32}\frac{1}{\eta^4}\left(\frac{\lambda}{a}\right)^4 \left(\frac{w}{\lambda}\right)^2,
\label{nesc}
\end{equation}
where $w$ is the waist of the incident Gaussian beam.  To obtain a large signal, the above expression suggests favorable scalings for small lattice spacing and large beam waist (within the array size). Moreover, the $\eta^{-4}\propto V$ scaling ($V$ being the trap depth), which  appears also in the average motion (\ref{av}), is favorable in the Lamb-Dicke regime, $\eta\ll 1$, that is assumed here. For example, for an illuminating beam of waist $w=4\lambda$, and taking the $a=0.5\lambda$ array parameters from Fig. 2, we find $n_{\mathrm{esc}}\approx27000$ photons, with a potential depth as low as $V=200E_R$ (recalling $n_{\mathrm{esc}}\propto V$).

\section{Discussion}
We developed an analytical formulation for the collective light-induced motion of atoms in a 2D array, relevant for typical optical-lattice systems in free space. This formalism can be seen as a generalization of the single-atom theory \cite{CCT}, to the case where resonant collective effects are significant, and taking advantage of the spatial lattice order of the atoms. We found the formation of light-induced collective mechanical modes of the array atoms, and analyzed their properties, such as the spectrum, stability and spatial structure, all of which determined by the incident light. These collective mechanical effects are predicted for atoms optically-trapped in free space, and are most significant for illumination close to the atomic resonance (more precisely, the cooperative dipolar resonance of the array). This is in contrast to previous studies of light-induced collective motion, which mainly focused on either atoms in confined geometries or on far off-resonant illumination \cite{GIO,OD2,CHA1,LIDDIna,RIT,RIT2,ESS,LEO}.

The above formalism should be useful for the study and estimation of atomic motion in optical lattices, especially in situations where its collective aspects, which are typically neglected, become important. This should be the case for light frequencies closer to resonance, and smaller lattice spacings. In particular, considering the recent theoretical studies of various resonant cooperative dipolar effects in 2D atomic arrays \cite{coop,ADM,janos,ADM2,janos2,CHA,HEN,ABA}, the estimation of these effects in any practical experimental scenario may require to consider the role of motion. As an example, we have estimated here the light-induced thermalization or``melting" rate of an atom array, and their effect on the observability of the reflected light.

In this respect, our formalism opens the way to the study of optomechanics of 2D atomic arrays. These arrays should exhibit strong nonlinear optical response induced by an optomechanical mechanism: Light that is incident on the  array induces the motion of its atoms, described by Eq. (\ref{90}), which in turn modifies the optical response of the array (e.g. its reflectivity) due to its strong dependence on the atomic positions and array geometry. Such a mechanism can lead to new quantum optomechanical effects.

\acknowledgments
We acknowledge fruitful discussions with Peter Zoller and Dominik Wild, and financial support from the NSF, the MIT-Harvard Center for Ultracold Atoms, and the Vannevar Bush Faculty Fellowship.

\appendix
\section{Induced dipole-dipole interactions}
In order to arrive at Eq. (\ref{HL}), we use the standard approach for the derivation of Heisenberg-Langevin equations in the Markov approximation (similar to Ref. \cite{LEH}, but including the atomic motion). We first write the Heisenberg equations of motion for the operators, $\hat{a}_{\mathbf{k}\mu}$, $\hat{\sigma}_{ni}$, $\hat{p}_n$ and $\hat{z}_n$. Solving formally for $\hat{a}_{\mathbf{k}\mu}(t)$ and inserting the solution into the equations for $\tilde{\sigma}_{ni}=\hat{\sigma}_{ni} e^{i\omega_L t}$ and $\hat{p}_n$, we take the Markov approximation; namely, we perform a coarse-graining in time, with a time resolution $\Delta t$ satisfying $|\mathbf{r}_{nm}|/c,1/\omega_L\ll \Delta t\ll \tau_s$, where $\tau_s$ is the fastest time-scale of atomic evolution, and $\mathbf{r}_{nm}=\mathbf{r}_{n}-\mathbf{r}_{m}$. We then need to evaluate the integrals
\begin{eqnarray}
\mathcal{D}_{ij}(\mathbf{r}_{nm})&=&\sum_{\mathbf{k}\mu}g^i_{\mathbf{k}\mu}(g^j_{\mathbf{k}\mu})^{\ast}e^{i\mathbf{k}\cdot\mathbf{r}_{nm}}\int_0^t dt'e^{-i(\omega_{\mathbf{k}}-\omega_L}(t-t'),
\nonumber\\
A_{ij}(\mathbf{r}_{nm})&=&\sum_{\mathbf{k}\mu}\hbar k_z g^i_{\mathbf{k}\mu}(g^j_{\mathbf{k}\mu})^{\ast}e^{i\mathbf{k}\cdot\mathbf{r}_{nm}}\int_0^t dt'e^{-i(\omega_{\mathbf{k}}-\omega_L}(t-t'),
\nonumber\\
\label{A1}
\end{eqnarray}
interpreted as the dipole-dipole interaction and force kernels, respectively, that appear as coefficients in the equations for $\tilde{\sigma}_{ni}$ and $\hat{p}_n$.

Within the Markov approximation, the dipole-dipole interaction kernel is found to be $\mathcal{D}_{ij}(\hat{\mathbf{r}}_{nm})\approx-i\frac{3}{2}\gamma\lambda G_{ij}(\omega_L,\hat{\mathbf{r}}_{nm})$ [Eq. (\ref{Dnm})], proportional to the dyadic Green's function at frequency $\omega=\omega_L=qc$,
\begin{equation}
G_{ij}(\mathbf{r})=\frac{e^{iqr}}{4\pi r}\left[\left(1+\frac{iqr-1}{q^2r^2}\right)\delta_{ij}+\left(-1+\frac{3-3iqr}{q^2r^2}\right)\frac{r^ir^j}{r^2}\right],
\label{G}
\end{equation}
with $r=|\mathbf{r}|$ and $r^i=\mathbf{e}_i\cdot\mathbf{r}$. For the force kernel, associated with laser-induced dipolar interactions, we note from (\ref{A1}) that $A_{ij}(\hat{\mathbf{r}}_{nm})=-i\hbar\left[\partial\mathcal{D}_{ij}(\mathbf{r})/\partial z \right]_{\mathbf{r}=\hat{\mathbf{r}}_{nm}}$ [Eq. (\ref{Anm})], obtaining
\begin{eqnarray}
&&\hat{A}^{ij}_{nm}\equiv A_{ij}(\hat{\mathbf{r}}_{nm})=-\frac{3}{4}\hbar q\gamma \left[\frac{\partial}{\partial
s_z}\overline{G}_{ij}(\mathbf{s})\right]_{\mathbf{s}=q(\hat{\mathbf{r}}_{nm})},
\nonumber\\
&&\frac{\partial}{\partial s_z}\overline{G}_{ij}(\mathbf{s})=\delta_{ij}\frac{s_z}{s}\frac{e^{is}}{s}
\nonumber\\
&&\times\left[\left(i-\frac{1}{s}\right)\left(1+\frac{is-1}{s^2}\right)+\left(\frac{i}{s^2}-2\frac{is-1}{s^3}\right)\right]
\nonumber\\
&&+\frac{s_i s_j s_z}{s^3}\frac{e^{is}}{s}
\nonumber\\
&&\times \left[\left(i-\frac{3}{s}\right)\left(-1+\frac{3-i3s}{s^2}\right)+3\left(-\frac{i}{s^2}-2\frac{1-is}{s^3}\right)\right]
\nonumber\\
&&+\left(\frac{s_j}{s}\delta_{iz}+\frac{s_i}{s}\delta_{jz}\right)\frac{e^{is}}{s}\left(-1+\frac{3-i3s}{s^2}\right)\frac{1}{s},
\label{A}
\end{eqnarray}
where $\overline{G}_{ij}(\mathbf{s})= 2\lambda G_{ij}(\mathbf{s}=q\mathbf{r})$ and $s_i=\mathbf{e}_i\cdot \mathbf{s}$.

For $i,j\neq z$, which is the relevant case for atoms that are polarizable only along the $xy$ plane, the last line of Eq. (\ref{A}) vanishes and we can write
\begin{eqnarray}
&&\hat{A}^{ij}_{nm}=-\frac{3}{4}\hbar q\gamma F_{ij}(q\hat{\mathbf{r}}_{nm})q(\hat{z}_n-\hat{z}_m),
\nonumber\\
&&F_{ij}(\mathbf{s})=\delta_{ij}\frac{e^{is}}{s^2}
\nonumber\\
&&\times\left[\left(i-\frac{1}{s}\right)\left(1+\frac{is-1}{s^2}\right)+\left(\frac{i}{s^2}-2\frac{is-1}{s^3}\right)\right]
\nonumber\\
&&+\frac{s_i s_j}{s^2}\frac{e^{is}}{s^2}
\nonumber\\
&&\times\left[\left(i-\frac{3}{s}\right)\left(-1+\frac{3-i3s}{s^2}\right)+3\left(-\frac{i}{s^2}-2\frac{1-is}{s^3}\right)\right].
\nonumber\\
\label{AF}
\end{eqnarray}
Considering the assumption $q |\hat{z}_n-\hat{z}_m|\ll 1$ from (\ref{46}), we further approximate
\begin{eqnarray}
\hat{A}^{ij}_{nm}&=&-\frac{3}{4}\hbar q\gamma F^{ij}_{nm}q(\hat{z}_n-\hat{z}_m),
\nonumber\\
F^{ij}_{nm}&=&F_{ij}(q\mathbf{r}^{\bot}_{nm}),
\label{F}
\end{eqnarray}
with $F_{ij}(\mathbf{s})$ from Eq. (\ref{AF}), such that $\hat{A}^{ij}_{nm}$ becomes linear in $(\hat{z}_n-\hat{z}_m)$. Therefore, the ``spring constant" $K_{nm}$ from Eq. (\ref{92}), which is proportional to $F^{ij}_{nm}$ [c.f. Eq. (\ref{92})], is proportional to the second derivative of the dipole-dipole interaction potential with respect to the translation $(\hat{z}_n-\hat{z}_m)$, as could be expected.

For two-level atoms, with dipole matrix element $\mathbf{d}=d\mathbf{e}_d$, $F_{ij}(\mathbf{s})$ from Eq. (\ref{AF}) simplifies to a scalar function,
$F_{nm}\equiv \mathbf{e}_d^{\dag}\cdot\overline{\overline{F}}_{nm}\cdot \mathbf{e}_d$,
as follows. In the first term, $\delta_{ij}$ gives $1$, whereas in the second term $s_i s_j$ depends on the specific polarization of the two-level transition, yielding for example $(s_x^2+s_y^2)/2$ for the case of circular polarization [$\mathbf{e}_d=(\mathbf{e}_x+i\mathbf{e}_y)/\sqrt{2}$], or $s_x^2$ for linear $x$ polarization [$\mathbf{e}_d=\mathbf{e}_x$]. For the examples that appear the figures, circular polarization is taken.

\begin{widetext}
\section{Coefficients and parameters of Eq. (\ref{90})}
For a general (paraxial) illumination from both sides of the array ($s=\pm$) we obtain

\begin{eqnarray}
&&\bar{f}_n=-\sum_{ss'=\pm}\hbar q\left[\frac{is\Omega_{ns}\Omega^{\ast}_{ns'}}{\delta_L-\Delta-i\frac{\gamma+\Gamma}{2}}+\mathrm{c.c.}\right],
\quad \quad \quad \quad \quad \quad\quad
\alpha_n=\frac{E_R}{\hbar}\sum_{ss'=\pm}\left[\frac{iss'\Omega_{ns}\Omega^{\ast}_{ns'}}{\left(\delta_L-\Delta-i\frac{\gamma+\Gamma}{2}\right)^2}+\mathrm{c.c.}\right],
\nonumber\\
&&\hat{f}_n(t)=\hbar q\sum_{ss'=\pm}\left[\frac{is\left(\Omega_{ns}\overline{\delta\Omega}_{ns'}^{\dag}+\delta\hat{\Omega}_{ns}\Omega_{ns'}^{\ast}\right)}{\delta_L-\Delta-i\frac{\gamma+\Gamma}{2}}+\mathrm{h.c.}\right],
\quad
K_{nm}=-\frac{3}{4}\hbar q^2\gamma\left[F_{nm}\sum_{ss'=\pm}\frac{\Omega_{ns}^{\ast}\Omega_{ms'}}{(\delta_L-\Delta)^2+\left(\frac{\gamma+\Gamma}{2}\right)^2}+\mathrm{c.c.}\right],
\nonumber\\
&&\hat{\alpha}_{nm}=
-\frac{3}{4}\frac{E_R}{\hbar}\frac{\gamma}{\delta_L-\Delta+i\frac{\gamma+\Gamma}{2}} F_{nm} q\hat{z}_{nm}
\sum_{ss'=\pm}\frac{s'\Omega_{ns}^{\ast}\Omega_{ms'}}{(\delta_L-\Delta)^2+\left(\frac{\gamma+\Gamma}{2}\right)^2},
\nonumber\\
&&\hat{f}_{nm}(t)=\frac{3}{4}\hbar q \gamma
\left[F_{nm}\sum_{ss'=\pm}\frac{q\hat{z}_{nm}\Omega^{\ast}_{ns}\overline{\delta\Omega}_{ms'}+\overline{\delta\Omega}^{\dag}_{ns}\Omega_{ms'}q\hat{z}_{nm}}{(\delta_L-\Delta)^2+\left(\frac{\gamma+\Gamma}{2}\right)^2} + \mathrm{h.c.} \right],
\label{90b}
\end{eqnarray}
where $\hat{z}_{nm}=\hat{z}_{n}-\hat{z}_{m}$ and $F_{nm}=\mathbf{e}_d^{\dag}\cdot\overline{\overline{F}}_{nm}\cdot \mathbf{e}_d$, with $F^{ij}_{nm}$ from Eq. (\ref{F}).
\end{widetext}

\section{Intuitive derivation of the collective force}
The collective force $K_{nm}(z_n-z_m)$ between a pair of atoms $n$ and $m$, can be intuitively derived as follows. The light-induced potential between a pair of two-level atoms linearly-driven by a far-detuned plane-wave laser is given by \cite{LIDDI}
\begin{equation}
U_{nm}=\frac{|\Omega|^2}{2\delta_L^2}\hbar \Delta_{nm}\cos(\mathbf{q}\cdot\mathbf{r}_{nm}),
\label{Unm}
\end{equation}
where $\Delta_{nm}$ is the resonant dipole-dipole interaction between the atoms, which is the imaginary part of $\mathcal{D}_{ii}(\mathbf{r}^{\bot}_{nm})$ considered above [c.f. Eq. (\ref{D})].
An intuitive way to understand this formula is to write it as $U_{nm}\propto I\alpha_n\alpha_m G_{nm}$, namely, the interaction is proportional to the intensity of the laser $I$, to the polarizabilities of atoms $\alpha_{n,m}$ (their linear response to the laser), and the photon Green's function between the atoms. In our case, Eq. (\ref{Unm}) has to be slightly modified as follows: (1) We allow to work on resonance, so $\delta_L^2$ in the denominator should be replaced by $\delta_L^2+(\gamma/2)^2$; (2) The linear responses of individual atoms have to be replaced by those of the array-renormalized atoms, leading to the replacements $\delta_L\rightarrow \delta_L-\Delta$ and $\gamma\rightarrow \Gamma+\gamma$; (3) In our case the laser is propagating normal to the array such that $\mathbf{q}\cdot \mathbf{r}_{nm}=0$ and the cosine gives $1$. Applying these replacements, the force between a pair of atoms along $z$ is found by differentiating $U_{nm}$ with respect to $z$, as in Eq. (\ref{Anm}), yielding the force $K_{nm}(z_n-z_m)$ from Eq. (\ref{92}). So, this force just originates from the laser-induced dipole-dipole potential $U_{nm}$ between the renormalized atoms $n$ and $m$.


\begin{thebibliography}{}
\bibitem{CCT} C. Cohen-Tannoudji, ``Atomic Motion in Laser Light", in \emph{Fundamental Systems in Quantum Optics}, Les Houches, Session LIII, 1990, pp. 1-164 (Elsevier Science Publisher B.V., 1992).
\bibitem{CCTb} C. Cohen-Tannoudji, J. Dupont-Roc, and G. Grynberg, \emph{Atom-Photon Interactions: Basic Processes and Applications}, (WILEY-VCH, 2004).
\bibitem{BEC1} M. H. Anderson, J. R. Ensher, M. R. Matthews, C. E. Wieman, and E. A. Cornell, Science \textbf{269}, 198 (1995).
\bibitem{BEC2} K. B. Davis, M.-O. Mewes, M. R. Andrews, N. J. van Druten, D. S. Durfee, D. M. Kurn, and W. Ketterle, Phys. Rev. Lett. \textbf{75}, 3969 (1995).
\bibitem{BEC3} C. C. Bradley, C. A. Sackett, J. J. Tollett, and R. G. Hulet, Phys. Rev. Lett. \textbf{75}, 1687 (1995).
\bibitem{PS} C. J. Pethick and H. Smith, \emph{Bose-Einstein Condensation in Dilute Gases} (Cabridge University Press, 2002).
\bibitem{OL} I.~Bloch, Nat. Phys. \textbf{1}, 23 (2005).
\bibitem{QS} I. Bloch, J. Dalibard and S. Nascimb\`{e}ne, Nat. Phys. \textbf{8}, pages 267 (2012).
\bibitem{TH} A. M. Kaufman, M. E. Tai, A. Lukin, M. N. Rispoli, R. Schittko, P. M. Preiss, M. Greiner, Science \textbf{353}, 794 (2016).
\bibitem{MBL} C. Gross and I. Bloch, Science \textbf{357}, 995 (2017).
\bibitem{QO} D. E. Chang, V. Vuletić, M. D. Lukin, Nat. Photon. \textbf{8}, 685 (2014).
\bibitem{RUO} J. Pellegrino, R. Bourgain, S. Jennewein, Y. R. P. Sortais, A. Browaeys, S. D. Jenkins, and J. Ruostekoski, Phys. Rev. Lett. \textbf{113}, 133602 (2014).
\bibitem{RUO2} S. D. Jenkins and J. Ruostekoski, Phys. Rev. A \textbf{86}, 031602(R) (2012).
\bibitem{coop} E. Shahmoon, D. Wild, M. Lukin and S. Yelin, Phys. Rev. Lett. \textbf{118}, 113601 (2017).
\bibitem{ADM} R. J. Bettles, S. A. Gardiner and C. S. Adams, Phys. Rev. Lett. \textbf{116}, 103602 (2016).
\bibitem{janos} J. Perczel, J. Borregaard, D. E. Chang, H. Pichler, S. F. Yelin, P. Zoller and M. D. Lukin, Phys. Rev. Lett. \textbf{119}, 023603 (2017).
\bibitem{ADM2} R. J. Bettles, J. Min\'{a}\v{r}, C. S. Adams, I. Lesanovsky and B. Olmos, Phys. Rev. A \textbf{96}, 041603(R) (2017).
\bibitem{janos2} J. Perczel, J. Borregaard, D. E. Chang, H. Pichler, S. F. Yelin, P. Zoller and M. D. Lukin, Phys. Rev. A \textbf{96}, 063801 (2017).
\bibitem{CHA} M. T. Manzoni, M. Moreno-Cardoner and A. Asenjo-Garcia, J. V. Porto, A. V. Gorshkov, and D. E. Chang, N. J. Phys. \textbf{20}, 083048 (2018).
\bibitem{HEN} L. Henriet, J. S. Douglas, D. E. Chang and A. Albrecht, arXiv:1808.01138 (2018).
\bibitem{ABA} V. Mkhitaryan, L. Meng, A. Marini, F. J. Garcia de Abajo, arXiv:1807.03231 (2018).
\bibitem{ANA} A. Asenjo-Garcia, M. Moreno-Cardoner, A. Albrecht, H. J. Kimble, and D. E. Chang, Phys. Rev. X \textbf{7}, 031024 (2017).
\bibitem{GIO} S. Giovanazzi, D. O'Dell and G. Kurizki, Phys. Rev. Lett. \textbf{88}, 130402 (2002).
\bibitem{OD2} D. H. J. O'Dell, S. Giovanazzi and G. Kurizki, Phys. Rev. Lett. \textbf{90}, 110402 (2002).
\bibitem{CHA1} D. E. Chang, J. I. Cirac and H. J. Kimble, Phys. Rev. Lett. \textbf{110}, 113606 (2013).
\bibitem{LIDDIna} E. Shahmoon, I. Mazets and G. Kurizki, Opt. Lett. \textbf{39}, 3674 (2014).
\bibitem{RIT} P. Domokos and H. Ritsch, Phys. Rev. Lett. \textbf{89}, 253003 (2002).
\bibitem{RIT2} T. Grie{\ss}er and H. Ritsch, Phys. Rev. Lett. \textbf{111}, 055702 (2013).
\bibitem{ESS} R. Mottl, F. Brennecke, K. Baumann, R. Landig, T. Donner and T. Esslinger, Science \textbf{336}, 1570 (2012).
\bibitem{LEO} J. L\'{e}onard, A. Morales, P. Zupancic, T. Esslinger and T. Donner, Nature \textbf{543}, 87 (2017).
\bibitem{NH} L. Novotny and B. Hecht,\emph{ Principles of Nano-Optics} (Cambridge University Press, 2006).
\bibitem{THI} T. Thirunamachandran, Mol. Phys. \textbf{40}, 393 (1980).
\bibitem{SAL} A. Salam, Advances in Quantum Chemistry \textbf{62}, 1 (2011).
\bibitem{LIDDI} E. Shahmoon and G. Kurizki, Phys. Rev. A \textbf{89}, 043419 (2014).
\bibitem{note2} E.g. for Rb87, where $\gamma\approx 2\pi\times 6$MHz and $\lambda\sim 780$nm, we find $\hbar\gamma/E_R\approx810$. Typical traping frequencies, $\nu_n$, are usually bounded by hundreds of kHz, validating also the condition $\tau_s^{-1}\gg \nu_n$, especially considering that $\tau_s^{-1}\sim \gamma+\Gamma$ can exceed $\gamma$ for $a<\lambda$ [Eq. (\ref{D})].
\bibitem{note3} E.g. for Rb87 we find that assumption (\ref{sp2}) corresponds to $v\ll 5$m/s (taking $\tau_s^{-1}\approx \gamma$), which seems rather unrestirctive for cold atoms, considering e.g. the $\sim 0.01$m/s thermal velocity of Rb87 at $1$ microkelvin.
\bibitem{note4} Considering $\omega_{\mathbf{k}}/c=k \sim q\pm 2\pi/(Tc)$ (due to $\delta^T_{\omega_{\mathbf{k}},\omega_L}$),
the correction to the approximaiton $e^{isk\hat{z}_n}\approx e^{isq\hat{z}_n}$ is a phase factor $e^{i\phi}$ with  $\phi\sim 2\pi \hat{z}_n/(Tc)=(\tau_s/T)(\tau_s\omega_L)^{-1}(\hat{z}_n/\lambda)4\pi^2$, which is expected to be extremely small and hence negligible: Its first two factors are guarented to be small by Eq. (\ref{T}) and $\tau_s^{-1}\ll \omega_L$, respectively, whereas the third factor should be smaller than unity for atoms contained in the traps.
\bibitem{note5} Upon performing the coarse-graining on the nonlinear term in the last line, we use the fact $\tilde{\sigma}_n$ relaxes to a steady-state right at the begining of the time-bin $T$ (separation of time scales, Eq. \ref{T}), and approximate
$\frac{1}{T}\int_{t_0}^{t_0+T}dt'\tilde{\sigma}_n^{\dag}(t')\tilde{\sigma}_m(t')\approx \frac{1}{T}\int_{t_0}^{t_0+T}dt' \overline{\sigma}_n^{\dag}(t)\overline{\sigma}_m(t)=\overline{\sigma}_n^{\dag}(t)\overline{\sigma}_m(t)$
\bibitem{QN} C. W. Gardiner and P. Zoller, \emph{Quantum Noise}, 2nd Edition (Springer-Verlag, Berlin Heidelberg, 2000).
\bibitem{DAL} J. Dalibard and C. Cohen-Tannoudji, J. Phys. B: At. Mol. Phys. \textbf{18}, 1661 (1985).
\bibitem{KRA} H.A. Kramers, Physica \textbf{7}, 284 (1940).
\bibitem{LEH} R. H. Lehmberg, Phys. Rev. A \textbf{2}, 883 (1970).


\end{thebibliography}
\end{document}